\documentclass[footinbib,a4paper,aps,prx,superscriptaddress,reprint,twocolumn,preprintnumbers,amsmath,amssymb,nobalancelastpage,10pt]{revtex4-1}
\usepackage{amsmath}
\usepackage{verbatim}
\usepackage{graphicx}
\usepackage{color}
\usepackage{tikz}
\usepackage{subfigure}
\usepackage{float}
\usepackage{slashed}
\usepackage{mathptmx}
\usepackage{amssymb}
\usepackage{amsmath}
\usepackage{amsfonts}
\usepackage{array}

\usepackage{bm}
\usepackage{xcolor}
\graphicspath{{figures/}}

\usepackage{braket}

\definecolor{mygrey}{gray}{0.35}
\definecolor{myblue}{rgb}{0.2,0.2,0.8}
\definecolor{mygreen}{rgb}{0.2,0.8,0.5}
\definecolor{myzard}{cmyk}{0,0,0.05,0}
\definecolor{mywhite}{rgb}{1,1,1}
\definecolor{myred}{rgb}{1,0.,0.3}

\usepackage[colorlinks=true,citecolor=myblue,linkcolor=myred]{hyperref}

 \def\ee{\mathord{\rm e}}
 
 \def\ii{\mathord{\rm i}}

\def\half{\textstyle\frac{1}{2}}

\renewcommand{\ii}{{\rm i}}
\renewcommand{\ee}{{\rm e}}
\def\beq{\begin{equation}}
\def\eeq{\end{equation}}
\def\barray{\begin{eqnarray}}
\def\earray{\end{eqnarray}}

\begin{document}

\title{  Large-$S$ and tensor-network methods for   strongly-interacting  topological insulators }


\author{E. Tirrito}
\affiliation{SISSA, Via Bonomea 265, 34136 Trieste, Italy}

\author{S. Hands}
\affiliation{Department of Mathematical Sciences, University of Liverpool, Liverpool L69 3BX, United Kingdom}

\author{A. Bermudez}
\affiliation{Instituto de F\'isica Te\'orica, UAM-CSIC, Universidad Autónoma de Madrid, Cantoblanco, 28049 Madrid, Spain. }

\begin{abstract}

The study of correlation effects in topological phases of matter can benefit from a multidisciplinary approach that combines techniques drawn from condensed matter, high-energy physics and quantum information science. In this work, we exploit these connections to study the strongly-interacting limit of certain lattice Hubbard models of topological insulators, which map onto four-Fermi quantum field theories with a Wilson-type discretization, and have been recently shown to be at reach of  cold-atom quantum simulators based on synthetic spin-orbit coupling. We combine large-$S$ and tensor-network techniques to explore the possible spontaneous symmetry-breaking  phases that appear when the interactions of the topological insulators are sufficiently large. In particular, we show that varying the Wilson parameter $r$ of the lattice discretizations leads to a novel Heisenberg-Ising compass model with critical lines that flow with the value of $r$. 


\end{abstract}

\maketitle

\setcounter{tocdepth}{2}
\begingroup
\hypersetup{linkcolor=black}
\tableofcontents
\endgroup

\section{\bf Introduction}

\subsection{Topological matter and  relativistic field theories}

Our most accurate description of Nature is based on a four-dimensional quantum field theory (QFT) of fermionic matter coupled to gauge fields: the standard model of particle physics~\cite{Peskin:1995ev}. In this context, current challenges arise in the understanding of effects  that cannot be treated perturbatively, such as  quark confinement  in quantum chromodynamics~\cite{greensite_2020}. To advance our understanding of non-perturbative phenomena, quantum field theorists have introduced other simplified models that share some important aspects  with the standard model but, at the same time, avoid the intricacies of non-Abelian gauge theories. Such toy QFTs, which are typically defined in reduced dimensionalities, have played an important role in elucidating phenomena such as asymptotic freedom, dynamical mass generation, chiral symmetry breaking, or the role of topological solutions and instantons.  Paradigmatic examples of such toy QFTs are the two-dimensional Thirring~\cite{THIRRING195891} and Gross-Neveu~\cite{PhysRevD.10.3235} models, which describe self-interacting Dirac fields,  and the two-dimensional non-linear sigma model~\cite{POLYAKOV197579}, which consists of scalar fields coupled through a non-linear constraint.  These models serve to develop and test tools, such as bosonization~\cite{PhysRevD.11.2088} and large-$N$ expansions~\cite{coleman_1985},  the predictions of which can be benchmarked with  efficient numerical methods  for such low-dimensional QFTs. Nonetheless, our most accurate experiments are consistent with a four-dimensional spacetime such that, in a strict sense,  the specific predictions of these  toy models are not solving  a specific  real  problems in high-energy physics that can be falsified  experimentally. Within high-energy physics, these toy QFTs  qualify instead as theoretical laboratories.  

Remarkably, during the last decades, we have witnessed a change of status for such low-dimensional  QFTs. Rather than looking at high energies and small length scales, one may instead focus on non-relativistic condensed-matter systems  which, at long wavelengths and small energies~\cite{WILSON197475}, display certain universal behaviour determined by  emergent relativistic QFTs. Such effective descriptions~\cite{Anderson393} provide a more flexible framework in comparison to the standard model, as realisation
properties such as the effective dimensionality or the emergent symmetries are not fixed a priori, but depend instead on the family of materials at hand. Some characteristic examples
where relativistic Dirac fields emerge include graphene~\cite{RevModPhys.81.109}, Weyl semimetals~\cite{RevModPhys.90.015001}, or topological insulators and superconductors~\cite{RevModPhys.83.1057}. For the emergence of  scalar fields subjected to non-linear constraints, quantum magnets play a prominent role~\cite{RevModPhys.63.1}.  

A further step in this direction is provided by the so-called quantum simulators (QSs)~\cite{Feynman_1982}: well-isolated quantum many-body systems with unparalleled levels of control down to the single-particle level that can directly mimic a specific target model~\cite{Cirac2012}. In the  context of low-dimensional QFTs,  QSs can be tailored such that one has full control of the microscopic parameters and, moreover, can access the continuum limit  in a controlled fashion. In order to do so, QSs of QFTs~\cite{doi:10.1002/andp.201300104,Zohar_2015,doi:10.1080/00107514.2016.1151199,Banuls2020,doi:10.1098/rsta.2021.0064,klco2021standard} typically follow the approach of lattice field theories~\cite{RevModPhys.51.659} in their Hamiltonian formulation~\cite{PhysRevD.11.395}. Rather than reducing the lattice spacing  to recover the continuum limit, one may tune the microscopic couplings of these QSs to approach a critical point where the correlation length is much larger than that spacing, and the continuum description sets in. In recent years, we have seen very promising  experimental steps in this direction for Dirac fermions~\cite{Gerritsma2010,PhysRevLett.106.060503} and Dirac QFTs~\cite{Tarruell2012,Duca288,Jotzu2014,PhysRevLett.121.150403, liang2021} and gauge theories~\cite{Martinez2016,z2LGT_2, Kokail2019,Mil1128,PhysRevX.10.021041,Yang2020} in low dimensions.

In this work, we explore the strong-coupling limit of four-Fermi models, namely QFTs of self-interacting Dirac fermions. As discussed below, these QFTs are  inspired by the Thirring and Gross-Neveu models, the origin of which can be traced back to  seminal contributions of E. Fermi~\cite{Fermi1934,doi:10.1119/1.1974382} and Y. Nambu and G. Jona-Lasinio~\cite{PhysRev.122.345,PhysRev.124.246}. In particular, we explore specific lattice discretizations based on the so-called Wilson fermions~\cite{Wilson1977}, which make direct connections of this four-Fermi QFTs  with the aforementioned topological insulators~\cite{RevModPhys.83.1057,PhysRevB.74.085308,PhysRevB.78.195424,Ryu_2010,PhysRevLett.105.190404,Mazza_2012,PhysRevLett.108.181807,Zache_2018}, allowing to study the effect of electron-electron  interactions. We note that recent advances in cold-atom QSs based on schemes of synthetic spin-orbit coupling in atomic gases with negligible interactions~\cite{PhysRevLett.121.150403, liang2021}  connect directly to these Wilson-regularised lattice field theories and, furthermore,    motivate a careful study of the regime of strong interactions. This is not only relevant from the perspective of QFTs, where one can find novel strongly-coupled fixed points that can only be characterised  non-perturbatively~\cite{HANDS199329,hep-lat/9706018}, but also from the perspective of strongly-correlated  effects in topological phases of matter, a topic that has received significant attention in recent years~\cite{Hohenadler_2013,Rachel_2018,Neupert_2015,doi:10.1142/S021797921330017X}. 

As we have discussed in a series of recent works, the native Hubbard interactions~\cite{PhysRevLett.81.3108,PhysRevLett.89.220407} of cold-atom QSs of spin-orbit coupling  in two~\cite{PhysRevX.7.031057,BERMUDEZ2018149,tirrito2021topological} and three~\cite{ziegler2020correlated,ziegler2021largen}  dimensions can be understood as the single-flavour-limit  of Four-Fermi QFTs with Lorentz-invariant self-interactions, and regularised via a Wilson-type discretization. Such a discretization introduces the Wilson parameter $r\in(0,1]$, which is customarily set to unity $r=1$ in most lattice studies. As briefly discussed in~\cite{ziegler2020correlated,ziegler2021largen}, setting $r<1$ has no important effect in the absence of interactions, as one can simply rescale the axes of the phase diagram in a simple manner to maintain the same layout: topological insulators are separated from trivial band insulators by critical lines in parameter space. The situation is not so clear as one switches on the interactions. Here, as a consequence of spontaneous symmetry breaking, one can find phases with long-range order  corresponding to different fermion condensates in the context of relativistic QFTs, as discussed in detail for $r=1$ ~\cite{BERMUDEZ2018149,tirrito2021topological,ziegler2020correlated,ziegler2021largen}. In this work, we explore the nature of these fermion condensates as the Wilson parameter is allowed to take values in $r<1$. To identify the possible condensates and chart the phase diagram, we explore the strong-coupling limit by deriving an effective  Heisenberg-type compass model with directional spin-spin interactions. Using the path-integral representation of the partition function,  we derive a version of the aforementioned non-linear sigma model with a discrete $\mathbb{Z}_2$ symmetry, a constrained QFT amenable to a large-$S$ expansion in the  limit where the effective spin $S\gg 1$. We then benchmark these predictions for the two- and three-dimensional lattice field theories with numerical simulations based on tensor networks.   

Besides the fundamental interest in understanding the role of the Wilson parameter in non-perturbative phenomena of four-Fermi models, we note that the specific cold-atom proposals  based on synthetic spin-orbit coupling~\cite{ziegler2020correlated,ziegler2021largen} lead to an effective Wilson parameter $R$ that is controlled by the ratio of spin-preserving and spin-flipping tunnelings, each of which can be independently controlled by the lasers that form a so-called optical Raman  lattice~\cite{PhysRevLett.121.150403, liang2021}. Accordingly, reaching the regime of $r=1$ would require additional fine tuning, as the generic QS would instead lead to $r\neq 1$. Regarding the possible experimental realisation of the four-fermi-Wilson model with cold atoms, it is thus an interesting and useful question to understand the effects of $0<r<1$.

\subsection{Constrained quantum field theories}

Let us start by discussing the nature of the constraints in representative QFTs, which will allow us to frame  the results of our work appropriately.
A well-known QFT where an effective constraint arises  is the $O(N)$ model, which describes a real scalar field $\boldsymbol{\Phi}(x)=(\phi_1(x),\cdots,\phi_N(x))^{\rm t}$ with $N$ flavours. In the    Hamiltonian formulation, and in the absence of interactions,  the free fields evolve under a Klein-Gordon Hamiltonian
\beq
\label{eq:H_0}
\mathcal{H}_0=\frac{1}{2}\left(\boldsymbol{\Pi}(x)\cdot \boldsymbol{\Pi}(x)-\partial^j\boldsymbol{\Phi}(x)\cdot\partial_j\boldsymbol{\Phi}(x)\right),
\eeq
where we use natural units $\hbar=c=1$, and Einstein's convention of repeated-index summation. Here, the fields and conjugate momenta $\boldsymbol{\Pi}(t,\boldsymbol{x})=\partial_t\boldsymbol{\Phi}(t,\boldsymbol{x})$ fulfil the canonical algebra $[\Phi_{\mathsf{f}_1}(t,\boldsymbol{x}_1),\Pi_{\mathsf{f}_2}(t,\boldsymbol{x}_2)]=\ii\delta_{\mathsf{f}_1,\mathsf{f}_2}\delta^d(\boldsymbol{x}_1-\boldsymbol{x}_2)$, and $j\in\{1,\cdots, d\}$ labels the spatial coordinates of a $D=(d+1)$-dimensional Minkowski spacetime with metric $\eta={\rm diag}(1,-1,\dots,-1)$. This QFT describes $N$ uncoupled scalar bosons, and is invariant under a continuous internal symmetry $\boldsymbol{\Phi}(x),\boldsymbol{\Pi}(x)\mapsto o\boldsymbol{\Phi}(x),o\boldsymbol{\Pi}(x)$, where $ o\in O(N)$ is an arbitrary rotation. In order to couple the different flavours, one introduces  a quartic self-interaction that respects this internal symmetry
\beq
\label{eq:int_H}
\mathcal{H}_{\rm int}=\frac{\lambda_0}{4!}\left(\boldsymbol{\Phi}(x)\cdot \boldsymbol{\Phi}(x)-\Phi_0^2\right)^2
\eeq
where $\lambda_0$ is the bare coupling strength. Here, we have introduced $\Phi_0$ as the vacuum expectation value attained by one of the scalar-field flavours, e.g. $\langle\Phi_\mathsf{f}(x)\rangle=\delta_{\mathsf{f},1}\Phi_0$, which corresponds to the spontaneous symmetry breaking (SSB) of the continuous $O(N)$ symmetry  in the classical limit, such that $O(N)\mapsto O(N-1)$. This corresponds to the meson sector of the linear sigma model~\cite{Gell-Mann1960}, which describes the coupling of $(N-2)(N+1)/2$ pions $\boldsymbol{\pi}$ to an additional heavy  scalar $\sigma$~\cite{Peskin:1995ev}, with $\sigma(x)=\Phi_1(x)/\Phi_0$, $\boldsymbol{\pi}(x)=(\Phi_2(x),\cdots\Phi_N(x))^{\rm t}/\Phi_0$, corresponding to the symmetry-breaking and Goldstone components respectively. Instead of expanding around the SSB groundstate, one may focus on the strong-coupling limit $\lambda_0\to\infty$, where the ground-state minimises the interaction energy~\eqref{eq:int_H} by imposing a non-linear constraint on the fields
\beq
\label{eq:bconstraint}
\sigma^2(x)+\boldsymbol{\pi}^2(x)=1,
\eeq
which are thus  forced to take values on the  unit sphere $S_{N-1}$. The Klein-Gordon field theory~\eqref{eq:H_0} subjected to this constraint~\eqref{eq:bconstraint} belongs to the family of sigma models~\cite{Gell-Mann1960}, which  describe particles forced to move on a specific manifold. In this particular case, this constrained model is called  the $O(N)$ {\it non-linear sigma model}~\cite{Peskin:1995ev}. As noted in the introduction, in two-dimensional spacetimes where the $O(N)$ symmetry cannot be spontaneously broken~\cite{Coleman1973}, the $O(N)$ non-linear sigma model shares important features with non-Abelian gauge theories such as asymptotic freedom for $N>2$~\cite{POLYAKOV197579}, existence of topologically non-trivial solutions called instantons~\cite{Polyakov1975MetastableSO}, or large-$N$ methods and dimensional transmutation~\cite{PhysRevD.14.985,coleman_1985}.

\begin{figure}[t]
	\centering
	\includegraphics[width=0.5\textwidth]{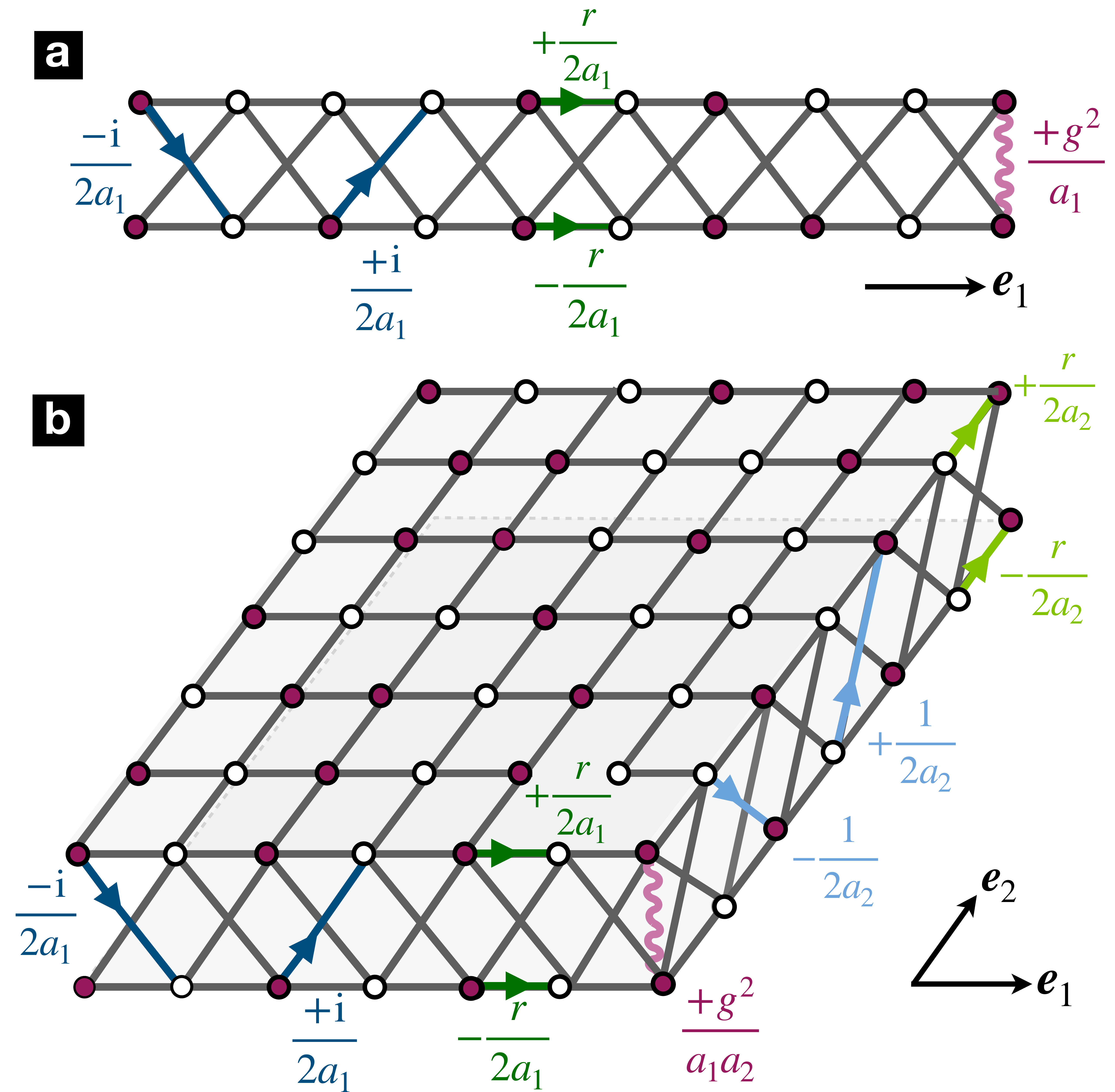}
	\caption{  {\bf Four-Fermi-Wilson field theories:} {\bf (a)} For $d=1$, the Dirac fermions indexes corresponding to lattice and spinor degrees of freedom can be depicted as a synthetic two-leg ladder. The Wilsonian regularisation of the four-Fermi field theory can be depicted by horizontal (intra-leg) and (cross-link inter-leg) tunnelings, as well as inter-leg density-density interactions, as well as an on-site energy imbalance (not shown).  {\bf (b)} For $d=2$, the fermions occupy a synthetic bilayer, with intra- and inter-layer tunnelings that depend on the direction $\{\boldsymbol{e}_j\}_{j=1,2}$, as well as an inter-layer density-density interaction, and an on-site energy imbalance (not shown). In both {\bf (a)} and {\bf (b)}, if the legs or layers are understood as different spin states of the fermions, the tunnelings can be understood in terms of spin-orbit coupling in Hubbard lattice models.  }
	\label{fig:fermion_model}
\end{figure}

 In this work, we will discuss how similar constraints can appear also in purely fermionic QFTs even in the absence of any continuous internal symmetry. In the most general setting, we consider a spinor field $\boldsymbol{\Psi}(x)=(\psi_1(x),\cdots,\psi_N(x))^{\rm t}$ with $N$ flavours  evolving under a Dirac Hamiltonian density
\beq
\label{eq:H_0_D}
\mathcal{H}_0=-\overline{\boldsymbol{\Psi}}(x)\ii(\mathbb{I}_N\otimes\gamma^j )\partial_j\boldsymbol{\Psi}(x),
\eeq
where we have introduced the gamma matrices $\{\gamma^\mu,\gamma^{\nu}\}=2\eta^{\mu\nu}$ for spacetime indexes $\mu,\nu\in\{0,1,\cdots, d\}$, and the adjoint  spinor $\overline{\boldsymbol{\Psi}}(x)={\boldsymbol{\Psi}}^\dagger(x)(\mathbb{I}_N\otimes\gamma^0)$. This model describes  $N$ uncoupled  Dirac fermions, and is invariant under a continuous unitary transformation $\boldsymbol{\Psi}(x)\mapsto u\otimes\mathbb{I}_{s}\boldsymbol{\Psi}(x)$, where $ u\in U(N)$, and the identity in the spinor components $\mathbb{I}_s$ depends on the dimensionality of the representation of the gamma matrices. Paralleling the  the discussion around  Eq.~\eqref{eq:int_H}, we can now couple the flavours via a four-Fermi~\cite{Fermi1934,doi:10.1119/1.1974382,PhysRev.122.345,PhysRev.124.246} contact interaction
\beq
\label{eq:int_H_D}
\mathcal{H}_{\rm int}=-\frac{g^2}{2N}\left(\overline{\boldsymbol{\Psi}}(x) \boldsymbol{\Psi}(x)\right)^2,
\eeq
where $g^2$ is the bare coupling strength.
Although not directly apparent, as in the bosonic case, we will show below that the strong-coupling limit $g^2\to\infty$ leads to a constraint similar to the one of Eq.~\eqref{eq:bconstraint}, where the $\sigma$ and $\boldsymbol{\pi}$ fields will be related to  particular SSB channels of the above  QFT related to  fermion condensates. In contrast to the bosonic case, the non-linear constraint appears down to the $N=1$ level, as the symmetry being broken is not the $U(N)$ symmetry, but rather a discrete $\mathbb{Z}_2$ symmetry involving the spinor degrees of freedom. In the following section, we will introduce a particular lattice discretization, which plays an important role in determining the  specific $\mathbb{Z}_2$ SSB, and its connection with topological insulators.

\subsection{ Four-Fermi interactions in topological insulators}

In this section, we describe in more detail the Wilson regularisation~\cite{Wilson1977} of the above fermionic QFT~\eqref{eq:H_0_D}-\eqref{eq:int_H_D}, and how it yields a playground to explore interactions in topological insulators. We consider the Hamiltonian lattice formulation~\cite{PhysRevD.11.395} obtained by   discretising  the spatial coordinates $\boldsymbol{x}\in\Lambda_d$,  focusing on the $d=1,2$ cases
\begin{equation}
\label{eq:lattices}
\Lambda_d=\Big\{\sum_{j=1}^dn_j a_j \boldsymbol{e}_j:\hspace{1ex}  n_j\in\mathbb{Z}_{N_{j}}\Big\},
\end{equation}
where $\{a_j\}$ are the lattice spacings along the $\{\boldsymbol{e}_j\}$ unit vectors, and $N_j$ are the corresponding number of lattice sites along each axis (see Fig.~\ref{fig:fermion_model}). Let us also note that  for $d=1,2$ spatial dimensions,  one can use the following irreducible representations of the gamma matrices
\beq
\label{eq:gamma_matrices}
\begin{split}
&d=1,\hspace{2ex} \gamma^0=\sigma_{\mathsf{z}},\hspace{1ex} \gamma^1=\ii\sigma_{\mathsf{y}},\\
&d=2,\hspace{2ex} \gamma^0=\sigma_{\mathsf{z}},\hspace{1ex} \gamma^1=\ii\sigma_{\mathsf{y}}, \hspace{1ex} \gamma^2=-\ii\sigma_{\mathsf{x}},
\end{split}
\eeq
which are proportional to the Pauli matrices.

Discretising the spatial derivatives appearing in  Eq.~\eqref{eq:H_0_D} using central differences leads to the so-called naive fermions~\cite{gattringer_lang_2010}, the continuum limit of which contains $N_{\rm D}=2^d$ Dirac fermions due to fermion doubling~\cite{NIELSEN198120,NIELSEN1981173}. We follow Wilson's prescription~\cite{Wilson1977}, which introduces additional terms that are responsible for giving different masses to each of these Dirac fermion species:
\begin{widetext} 
\beq \label{eq:H_lattice_Wilson}
\mathcal{H}=\sum_{j=1}^d\!\!\left(\!-\overline{\boldsymbol{\Psi}}(\boldsymbol{x})\left(\frac{\ii(\mathbb{I}_N\otimes\gamma^j)}{2a_j}+\frac{r{(\mathbb{I}_N\otimes\mathbb{I}_s)}}{2a_j}\right)\boldsymbol{\Psi}(\boldsymbol{x}+a_j\textbf{e}_j)+\overline{\boldsymbol{\Psi}}(\boldsymbol{x})\left(\frac{m}{2d}+\frac{r}{2a_j}\right)\boldsymbol{\Psi}(\boldsymbol{x})+{\rm
H.c.}\right)-\frac{g^2}{2N}\bigg(\!\overline{\boldsymbol{\Psi}}(\boldsymbol{x}){\boldsymbol{\Psi}}(\boldsymbol{x})\!\!\bigg)^{\!\!2},
\eeq 
\end{widetext} 
where we have introduced the bare mass $m$, and the aforementioned dimensionless Wilson parameter $r$. In Fig.~\ref{fig:fermion_model}, we present a schematic diagram of this Wilsonian discretization by means of tunnelling processes and density-density interactions with strengths obtained after rescaling the fields in terms  of dimensionless creation-annihilation operators. In lattice field theories (LFTs), one typically works directly with the Euclidean action associated to the above Hamiltonian by also discretising the Wick-rotated temporal coordinate $x=\sum_{\mu=0}^d a_\mu n_\mu\boldsymbol{e}_\mu$, such that recovering the time-continuum limit requires temporal anisotropies that permit the limit  $a_0\to 0$. In case one is not interested in making contact with the Hamiltonian formulation, it is possible to focus directly on the isotropic regime $a_\mu=a$, and consider $|r|\leq 1$,  as imposed by the reflection positivity of the Euclidean action for $g^2=0$
~\cite{montvay_munster_1994}. A standard choice in the literature is to set $r=1$, such that  one recovers a single massless Dirac fermion in the limit of $m\to 0$ and at long wavelengths $a\to 0$, while the remaining doublers acquire a very large mass proportional to $1/a$, and thus lie at the UV cutoff of the regularised QFT. The choice $r=1$ brings the technical advantage that tunnelling terms are proportional to projection operators $P_{j\pm}\equiv{1\over2}(1\pm\ii\gamma^j)$ with $P^2_{j\pm}=P_{j\pm}$, $P_{j\pm}P_{j\mp}=0$.

The goal of the present work is to explore regimes with $0<r<1$, and make connections with effective constrained QFTs  in the strong-coupling limit. Likewise, we will also explore  anisotropic lattice constants $a_\mu\neq a_\nu$. We remark that  isotropy is not required {\it a priori}, since  the continuum limit yields a QFT invariant under the full Lorentz group $SO(1,d)$ even when the anisotropic lattice formulation breaks translational, rotational and Lorentz symmetries explicitly. In fact,  temporal~\cite{PhysRevLett.111.172001} and spatial anisotropies~\cite{_CLQCD__2007} can actually increase the accuracy of  lattice computations. To understand the effect of non-unity Wilson parameters and anisotropic lattice constants, we can start by focusing on the non-interacting limit $g^2=0$. In this case, it is straightforward to compute the half-filled groundstates $\ket{\epsilon(\boldsymbol{k})}$ corresponding to the Dirac vacua~\cite{BERMUDEZ2018149,ziegler2020correlated,ziegler2021largen}, where we have introduced the quasi-momentum within the first Brillouin zone $\boldsymbol{k}\in\mathsf{BZ}=\times_j \left(-{\pi}/{a_j},{\pi}/{a_j}\right]$. Associated to this band structure, one finds a Berry connection $\boldsymbol{\mathcal{A}}(\boldsymbol{k})=\bra{\epsilon(\boldsymbol{k})}\ii\boldsymbol{\nabla}_{\boldsymbol{k}}\ket{\epsilon(\boldsymbol{k})}$~\cite{doi:10.1098/rspa.1984.0023, berry_review}
, which characterises the principal fibre bundle associated to the  occupied energy band~\cite{nakahara_2017}. Such fibre bundles can be characterised by topological invariants which depend on dimensionality.

For $d=1$ spatial dimensions, Zak's phase~\cite{PhysRevLett.62.2747} allows to define the  Wilson loop for a cycle in momentum space
\beq
\label{eq:zak}
W_{\rm Z}=\ee^{\ii\varphi_{\rm Z}},\hspace{1ex}\varphi_{\rm Z}=\!\!\int\!\!{\rm d}k\mathcal{A}(\boldsymbol{k})= N\pi\left(\theta(2r+ma_1)-\theta(ma_1)\right),
\eeq
where $\theta(x)$ is the Heaviside step function. Therefore, 
one finds   a trivial band insulator with $W_{\rm Z}=+1$ for $ma_1>0$ or $ma_1<-2r$. Alternatively, a non-trivial topological insulator $W_{\rm Z}=-1$ arises when $-2r<ma_1<0$ and the number of flavours $N$ is odd, which actually lies  in the symmetry class $\mathsf{BDI}$~\cite{PhysRevB.78.195125,doi:10.1063/1.3149495}.
In comparison to the previous results of~\cite{BERMUDEZ2018149}, which focused on the standard choice $r=1$, we observe that the structure of  the non-interacting phase diagram is completely equivalent if one simply rescales the dimensionless mass with the Wilson parameter $ma_1\mapsto m a_1/r$.

For $d=2$, the Berry curvature ${\mathcal{B}}(\boldsymbol{k})=\boldsymbol{\nabla}_{\boldsymbol{k}}\wedge\boldsymbol{\mathcal{A}}(\boldsymbol{k})$~\cite{doi:10.1098/rspa.1984.0023} allows to define the first Chern number 
\beq
\label{eq:chern}
\begin{split}
N_{\rm Ch}=\!\!\int\!\!\frac{{\rm d}^2k}{2\pi}\mathcal{B}(\boldsymbol{k})= &N\left(\theta(2r\xi_2+ma_1)-\theta(ma_1)\right.\\
&\left.+\theta(2r(1+\xi_2)+ma_1)-\theta(2r+ma_1)\right).
\end{split}
\eeq
where $\xi_2=a_1/a_2$ is an anisotropy ratio, and we have assumed $\xi_2\leq 1$. Here, some comments are in order. In the isotropic limit $\xi_2=1$, and setting $r=1$, the groundstate corresponds to quantum anomalous Hall ($\mathsf{QAH}$) phase with $N_{\rm Ch}=-N$ for $-2<ma_1<0$, and $N_{\rm Ch}=+N$ for $-4<ma_1<-2$, whereas it is a trivial band insulator with $N_{\rm Ch}=0$ for $ma_1>0$ or $ma_1<-4$. This limit can be readily mapped onto the  Qi-Wu-Zhang model~\cite{PhysRevB.74.085308,PhysRevB.78.195424} of the $\mathsf{QAH}$~\cite{doi:10.1146/annurev-conmatphys-031115-011417}, with a central region along the $ma_1$ axis comprising the $\mathsf{QAH}$ phases, surrounded by trivial band insulators at both sides. It is worth noting that both the $\mathsf{QAH}$ and the $\mathsf{BDI}$ topological insulators can be understood as the bulk of a lower-dimensional version of the domain-wall-fermion construction of  lattice field theories~\cite{KAPLAN1992342,Kaplan:2021ewi}, in which the non-zero topological invariants would give rise to effective Chern-Simons-type terms in the response of the fermions to external gauge fields~\cite{GOLTERMAN1993219,PhysRevB.78.195424}.

As discussed in~\cite{ziegler2020correlated,ziegler2021largen,PhysRevD.102.094520}, allowing for spatial anisotropies $\xi_2<1$, and fixing the Wilson parameter to the standard value $r=1$, one  finds an additional trivial band insulator for $-2<ma_1<-2\xi_2$ separating the two QAH phases with $N_{\rm Ch}=\pm N$. Something completely analogous occurs for spatial anisotropies  $\xi_2>1$. From the above expression~\eqref{eq:chern}, we see again that the effect of a non-unity Wilson parameter $0<r<1$ is rather trivial in the non-interacting case, one can simply rescale $ma_1\mapsto m a_1/r$, and obtain the same structure and phases as in the limit of $r=1$. However, as one switches on interactions $g^2>0$, the situation need not be so simple: there can be SSB processes that lead to long-range-ordered phases different from the above trivial and topological insulators. In the following sections, we will extend our previous studies in~\cite{BERMUDEZ2018149,ziegler2020correlated,ziegler2021largen},  studying the nature of these SSB processes for arbitrary Wilson parameters $0<r<1$ by exploring the strong-coupling limit, in which the four-Fermi interaction strength is the leading parameter $g^2\to\infty$. As discussed below, a different constrained QFT controls that  limit, which will be exploited to predict the shape of the phase diagram and possible phase transitions.

\section{\bf Strong couplings and  effective spin models}
\label{sec:eff_spin_model}

\subsection{ Ising order and fermion condensates}

\begin{figure}[t]
	\centering
	\includegraphics[width=0.45\textwidth]{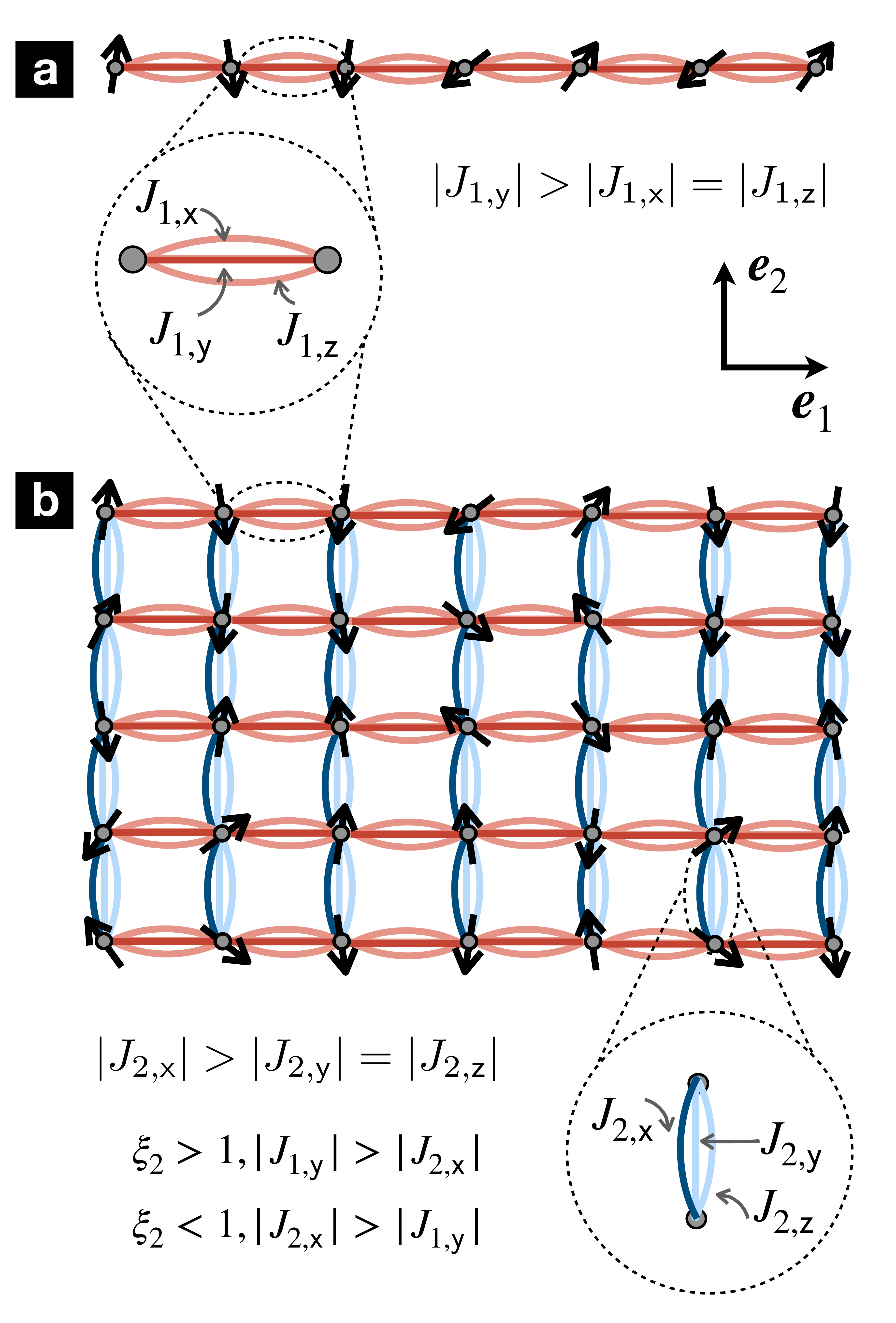}
	\caption{  {\bf Strong-coupling Heisenberg-Ising spin models:} {\bf (a)} For $d=1$, the half-filled chain $\Lambda_1$ in the limit of strong interactions can be described by localised spins $S=1/2$, here depicted with black arrows.  The spins interact with the neighbours via spin -spin couplings $S_{\mathsf{a}}(\boldsymbol{x})S_{\mathsf{a}}(\boldsymbol{x}+a_1\boldsymbol{e}_1)$ of strength $J_{1,\mathsf{a}}$ for $\mathsf{a}\in\{\mathsf{x,y,z}\}$, here represented by solid lines in a red scale that determines their relative magnitude for a generic Wilson parameter $0<r<1$, such that $|J_{1,\mathsf{y}}|$ dominates. {\bf (b)} For the $d=2$ half-filled lattice $\Lambda_2$, the effective spins $S=1/2$ of the strong-coupling limit are arranged in a rectangular lattice, and  the spin-spin couplings $S_{\mathsf{a}}(\boldsymbol{x})S_{\mathsf{a}}(\boldsymbol{x}+a_j\boldsymbol{e}_j)$ have a directional character $J_{j,\mathsf{a}}$, where $j\in\{1,2\}$ labels the horizontal and vertical neighbours, leading to a compass-type model. In addition to the horizontal interactions in {\bf (a)}, the spins now interact vertically with strengths $J_{2,\mathsf{a}}$, here represented by solid lines in a blue scale that determines their relative magnitude for $0<r<1$, such that $|J_{2,\mathsf{x}}|$ dominates. The anisotropy parameter $\xi_2=a_1/a_2$ controls the directionality of the compass Heisenberg-Ising model, i.e. if the vertical ($|J_{1,\mathsf{a}}|<|J_{2,\mathsf{a}}|)$ or horizontal ($|J_{1,\mathsf{a}}|>|J_{2,\mathsf{a}}|$) couplings dominate, which occurs for $\xi_2<1$ and $\xi_2>1$, respectively.  }
	\label{fig:spin_model}
\end{figure}

In order to understand the strong-coupling limit, let us first note that for $d=1,2$ spatial dimensions,  the  irreducible representations of the gamma matrices~\eqref{eq:gamma_matrices}
imply that the Dirac spinors have two components $\psi_{\mathsf{f}}(x)=(\psi_{\mathsf{f},1}(x),\psi_{\mathsf{f},2}(x))$. In the  single-flavour limit $\mathsf{f}=1=N$, and for a fixed total number of fermions, the four-Fermi term in Eq.~\eqref{eq:H_lattice_Wilson} can be rewritten as
\beq
\mathcal{H}_{\rm int}=g^2\psi^\dagger_{\mathsf{f},1}(x)\psi^{\phantom{\dagger}}_{\mathsf{f},1}(x)\psi^\dagger_{\mathsf{f},2}(x)\psi^{\phantom{\dagger}}_{\mathsf{f},2}(x),
\eeq
up to an irrelevant shift of the ground-state energy. Accordingly, 
 the strong-coupling limit $g^2\to\infty$ will give rise to a large energy penalty for  configurations in which a pair of fermions occupy the same lattice site.  The Dirac vacuum corresponding to the half-filled groundstate  will have  a single fermion per site $\boldsymbol{n}$, which has the freedom to  select one  of the two possible spinor configurations $\ket{\uparrow_{\boldsymbol{n}}},\ket{\downarrow_{\boldsymbol{n}}}$. Within this subspace, the operators that fulfil the $SU(2)$ algebra $[\hat{S}_{\mathsf{a}}({t,\boldsymbol{x}}),\hat{S}_{\mathsf{b}}(t,\boldsymbol{x}')]=\ii\delta_{\boldsymbol{n},\boldsymbol{n}'}\sum_{\mathsf{c}}\epsilon_{\mathsf{abc}}\hat{S}_{\mathsf{c}}({t,\boldsymbol{x}})$ at spatial points $\boldsymbol{x}=\sum_jn_ja_j\boldsymbol{e}_j$, $\boldsymbol{x}'=\sum_jn'_ja_j\boldsymbol{e}_j$, become
\beq
\label{eq:spin_operators}
\boldsymbol{\hat{S}}({x})=S\Big(\prod_ja_j\Big)\psi^\dagger_{\mathsf{f}}(x)\boldsymbol{\sigma}\psi^{\phantom{\dagger}}_{\mathsf{f}}(x)\mapsto \boldsymbol{S}({x})=S\boldsymbol{\sigma}_{\boldsymbol{n}}.
\eeq
Here,  $S=1/2$, and $\boldsymbol{\sigma}_{\boldsymbol{n}}$  is an operator acting on the projected Hilbert space $\mathsf{H}_{\rm eff}=\otimes_{\boldsymbol{n'}}{\rm span}\{\ket{\uparrow_{\boldsymbol{n'}}},\ket{\downarrow_{\boldsymbol{n'}}}\}$, and defined by the tensor product of  the identity $\mathbb{I}_2$ on all sites except for $\boldsymbol{n}$, where one applies the  vector of Pauli matrices $\boldsymbol{\sigma}=(\sigma_{\mathsf{x}},\sigma_{\mathsf{y}},\sigma_{\mathsf{z}})$. 
As discussed in~\cite{BERMUDEZ2018149,ziegler2020correlated,ziegler2021largen} for $d=1,2$, and for unit Wilson parameter $r=1$, the lattice model that controls this strong-coupling limit corresponds to an effective spin model, where the spins reside on the sites of the spatial lattice regularisation~\eqref{eq:lattices}. These spins are subjected to local on-site terms, and interact with each other via  nearest-neighbour couplings depicted in Fig.~\ref{fig:spin_model}. The physical mechanism underlying these nearest-neighbour spin-spin interactions is the so-called super-exchange~\cite{PhysRev.79.350,ANDERSON196399}, and the most-general effective Hamiltonian can be written as follows
\beq
\label{eq:spin_model}
H_{\rm eff}=\sum_{\boldsymbol{x}\in\Lambda_d}\sum_{\mathsf{a}}\left(\sum_{j=1}^dJ_{j,\mathsf{a}}S_\mathsf{a}(\boldsymbol{x})S_\mathsf{a}(\boldsymbol{x}+a_j\boldsymbol{e}_j)+h_{\mathsf{a}} S_{\mathsf{a}}(\boldsymbol{x})\right).
\eeq 
Here,  we have used the label $\mathsf{a}\in\{\mathsf{x,y,z}\}$ to distinguish the internal spin components   from  the spatial coordinates $j\in\{1,\cdots,d\}$, and introduced a set of spin-spin couplings $J_{j,\mathsf{a}}$ describing the strength of the  interactions between neighbouring spins connected by a $\boldsymbol{e}_j$ link, and coupling their  internal   spin components via a $S_\mathsf{a}S_\mathsf{a}$ interaction (see Fig.~\ref{fig:spin_model}).

In previous works~\cite{BERMUDEZ2018149,ziegler2020correlated,ziegler2021largen} where we used the standard choice $r=1$, the nature of the spin-spin couplings was restricted to be of Ising type. For $d=1$, where the coupling strength  $g^2$ is dimensionless, the   spin couplings found were
\beq
\label{eq:couplings_1d}
J_{1,\mathsf{a}}=-\frac{2}{g^2a_1}\delta_{\mathsf{a},\mathsf{y}},\hspace{1ex} h_{\mathsf{a}}=2\left(m+\frac{1}{a_1}\right)\delta_{\mathsf{a},\mathsf{z}},
\eeq
which have units of inverse length, such that the effective Hamiltonian~\eqref{eq:spin_model} with dimensionless spin operators~\eqref{eq:spin_operators} has the correct units of energy. The 
strong-coupling Hamiltonian for $r=1$ thus coincides with a quantum Ising model~\cite{PFEUTY197079,RevModPhys.51.659} with $S_{\mathsf{y}}S_{\mathsf{y}}$ ferromagnetic interactions subjected to a transverse field along the internal $\mathsf{z}$  axis~\cite{BERMUDEZ2018149}. This is an exactly-solvable model with a quantum phase transition at $|h_{\mathsf{z}}|=|J_{1,\mathsf{y}}|/2$, marking the onset of SSB of an underlying  $\mathbb{Z}_2$ symmetry $\boldsymbol{S}(\boldsymbol{x})\mapsto(-{S}_{\mathsf{x}}(\boldsymbol{x}),-{S}_{\mathsf{y}}(\boldsymbol{x}),{S}_{\mathsf{z}}(\boldsymbol{x}))$, $\forall\boldsymbol{x}\in\Lambda_1$. This symmetry can also be combined with a reflection about the lattice centre,  such that 
\beq
\label{eq:symmetry_spin_model}
\boldsymbol{S}(\boldsymbol{x})\mapsto(-{S}_{\mathsf{x}}(-\boldsymbol{x}),-{S}_{\mathsf{y}}(-\boldsymbol{x}),{S}_{\mathsf{z}}(-\boldsymbol{x})).
\eeq
 It is interesting to note that, for the representation of the gamma matrices in 
Eq.~\eqref{eq:gamma_matrices}, this $\mathbb{Z}_2$ symmetry corresponds to 
\beq
\label{eq:inversion_symmetry}
\psi_{\mathsf{f}}(t,\boldsymbol{x})\mapsto\gamma^0\psi_{\mathsf{f}}(t,-\boldsymbol{x}),
\eeq
which is precisely the parity symmetry on Dirac spinors~\cite{Peskin:1995ev}. Accordingly, the ferromagnet with all spins aligned with the internal $\mathsf{y}$ axis (FM$_{\mathsf{y}}$) can be readily identified with a parity-breaking pseudo-scalar $\pi$ condensate 
\beq
\label{eq:pseudo-sclalr_condensate}
\langle S_{\mathsf{y}}(\boldsymbol{x})\rangle\propto\Pi_5=\langle\overline{\psi}_{\mathsf{f}}(\boldsymbol{x})\ii\gamma^5{\psi}_{\mathsf{f}}(\boldsymbol{x})\rangle,
\eeq
  where $\gamma^5=\gamma^0\gamma^1$ is the chiral gamma matrix. Note that, in the quantum Ising model~\cite{PFEUTY197079}, the groundstate always displays a non-zero magnetisation along the direction of the transverse field for any $h_{\mathsf{z}}\neq 0$. In the language of Dirac spinors, this corresponds to a non-zero value of the so-called scalar $\sigma$ condensate
 \beq
\langle S_{\mathsf{z}}(\boldsymbol{x})\rangle\propto\Sigma=\langle\overline{\psi}_{\mathsf{f}}(\boldsymbol{x}){\psi}_{\mathsf{f}}(\boldsymbol{x})\rangle,
 \eeq
 and the fact that is generically non-zero can be traced back to the explicit breaking of the discrete chiral symmetry $
\psi_{\mathsf{f}}(x)\mapsto\gamma^5\psi_{\mathsf{f}}(x),
$ by the Wilson discretisation~\eqref{eq:H_lattice_Wilson}. The appearance of the scalar condensate is typical of four-Fermi models with dynamical mass generation, such as the Gross-Neveu model~\cite{PhysRevD.10.3235}, whereas the pseudo-scalar one depends on the specific lattice regularisation. A non-zero pseudo-scalar condensate has also been discussed in the context of  lattice gauge theories in 3+1 dimensions~\cite{PhysRevD.30.2653,PhysRevD.58.074501}, and is known as the Aoki phase. We thus see that the strong-coupling limit captures nicely this condensate even in the $N=1$ limit and that, moreover, it provides an analytical expression for the critical line that was proved to be very accurate by comparing with matrix-product-state numerics~\cite{BERMUDEZ2018149}. In this manuscript, we will explore how this situation changes as the Wilson parameter is modified $0<r<1$.

For $d=2$ spatial dimensions,  where the coupling strength $g^2$ has units of length,  setting $r=1$~\cite{ziegler2020correlated,ziegler2021largen} leads to the   following spin-spin couplings 
\beq
\label{eq:couplings_2d}
 J_{1,\mathsf{a}}\!=\!\frac{-2a_2}{g^2a_1}\delta_{\mathsf{a},\mathsf{y}},\hspace{0.25ex}J_{2,\mathsf{a}}\!=\!\frac{-2a_1}{g^2a_2}\delta_{\mathsf{a},\mathsf{x}},\hspace{0.25ex} h_{\mathsf{a}}\!=2\!\!\left(\!\!m+\frac{1}{a_1}+\frac{1}{a_2}\!\right)\!\!\delta_{\mathsf{a},\mathsf{z}},
\eeq
which again have the correct units of energy as in  Eq.~\eqref{eq:couplings_1d}. The corresponding spin model~\eqref{eq:spin_model} is an instance of the so-called 90$^{\rm o}$ compass model~\cite{PhysRevB.71.195120,RevModPhys.87.1} with directional spin couplings $S_{\mathsf{y}}S_{\mathsf{y}}$ ($S_{\mathsf{x}}S_{\mathsf{x}}$) along the $\boldsymbol{e}_1$ ($\boldsymbol{e}_2$) spatial axis, and a transverse magnetic field again directed along the  internal $\mathsf{z}$ axis. In contrast to the quantum Ising chain~\eqref{eq:couplings_1d}, the compass model is no longer exactly solvable, and presents two different types  of phase transition. For $h_{\mathsf{z}}=0$, which is achieved for a negative bare mass $m=-1/a_1-1/a_2$, there is a well-studied  first-order phase transition at  $J_{1,\mathsf{y}}= J_{2,\mathsf{x}}$
~\cite{dorier2005quantum,chen2007quantum,orus2009first}. This critical point separates two different  ferromagnets: a FM$_{\mathsf{x}}$ characterised by the order parameter $|\langle S_{\mathsf{x}}(\boldsymbol{x})\rangle|>0$ for $|J_{2,\mathsf{x}}|>|J_{1,\mathsf{y}}|$, achieved for $a_1>a_2$, and a FM$_{\mathsf{y}}$  characterised by  $|\langle S_{\mathsf{y}}(\boldsymbol{x})\rangle|>0$ for $|J_{1,\mathsf{y}}|>|J_{2,\mathsf{x}}|$ for $a_2>a_1$. Both ferromagnets break the aforementioned $\mathbb{Z}_2$ symmetry~\eqref{eq:symmetry_spin_model}. We remark that in this $d=2$ case, the expression of this symmetry in fermion operators~\eqref{eq:inversion_symmetry} does not correspond to parity, as ${\bf x}\mapsto-{\bf x}$ is generated by a rotation in the connected component of the Lorentz group $SO(1,2)$. Rather than breaking parity,  a non-zero value of the corresponding fermion $\pi$ condensates  
\beq
\label{eq:pi_condensates}
\begin{split}
\langle S_{\mathsf{x}}(\boldsymbol{x})\rangle\propto\Pi_1=\langle\overline{\psi}_{\mathsf{f}}(\boldsymbol{x})\gamma^1{\psi}_{\mathsf{f}}(\boldsymbol{x})\rangle,\\
 \langle S_{\mathsf{y}}(\boldsymbol{x})\rangle\propto\Pi_2=\langle\overline{\psi}_{\mathsf{f}}(\boldsymbol{x})\gamma^2{\psi}_{\mathsf{f}}(\boldsymbol{x})\rangle,
 \end{split}
\eeq  breaks inversion symmetry. We note that taking a continuum  long-wavelength limit around the critical lines that separate these ferromagnets from the symmetric paramagnet would lead to a QFT where Lorentz symmetry cannot be recovered when approaching from the condensed phase. Accordingly, these $d=2$ FM$_{\mathsf{x}}$, FM$_{\mathsf{y}}$ phases were referred in~\cite{ziegler2020correlated,ziegler2021largen} as Lorentz-breaking condensates, which contrast the parity-breaking pseudo-scalar condensate~\eqref{eq:pseudo-sclalr_condensate} of $d=1$.

In contrast, the regime with a non-vanishing transverse field $h_{\mathsf{z}}\neq0$ has not been studied in so much detail. In the limit of very large spatial anisotropies $\xi_2=a_1/a_2\to 0$ ($\xi_2=a_1/a_2\to \infty$), the compass model~\eqref{eq:couplings_2d} reduces to a  collection of uncoupled rows (columns), each described by an Ising model in a transverse field with a second-order phase transition  at $|h_{\mathsf{z}}|=|J_{1,\mathsf{y}}|/2$ ($|h_\mathsf{z}|=|J_{2,\mathsf{x}}|/2$). This critical point separates a paramagnet, which has all spins  aligned along the internal $\mathsf{z}$ axis, from the aforementioned ferromagnet with  $|\langle S_{\mathsf{y}}(\boldsymbol{x})\rangle|>0$ ( $|\langle S_{\mathsf{x}}(\boldsymbol{x})\rangle|>0$) for each row (column). We note that there is an accidental exponentially-large degeneracy in the number of rows (columns) in these large-anisotropy limits. For  non-zero anisotropies $\xi_2>0$, yet finite $\xi_2^{-1}\neq 0$, these rows (columns) become coupled, lifting the degeneracy and selecting a unique 2-fold degenerate ferromagnetic ground-state FM$_{\mathsf{x}}$ (FM$_{\mathsf{y}}$) where all columns (rows) get locked to the same  spin direction  when  $|h_{\mathsf{z}}|<|J_{2,\mathsf{x}}|/\zeta$ and $|J_{2,\mathsf{x}}|>|J_{1,\mathsf{y}}|$  ($|h_{\mathsf{z}}|<|J_{1,\mathsf{y}}|/\zeta$  and $|J_{1,\mathsf{y}}|>|J_{2,\mathsf{x}}|$). Here, we have introduced a parameter $zeta$, which serves to locate the critical point, and will be equal to $\zeta=2$ in the regime of large spatial anisotropies. For other finite and non-zero anisotropies $\xi_2=a_1/a_2$, the critical point will change, and one should find that $\zeta\neq2$, which can no longer be found exactly, but must be estimated using numerical or analytical approximations~\cite{ziegler2020correlated,ziegler2021largen}.

\subsection{ Heisenberg-Ising chains for $d=1$}
\label{sec:Heis_d_1}

So far, we have reviewed  known results that  apply for Wilson parameter $r=1$.  Moving away from this limit modifies the super-exchange mechanism, leading to additional spin-spin couplings depicted
in Fig.~\ref{fig:spin_model} {\bf (a)}. These still admit the general form of Eq.~\eqref{eq:spin_model}, but lead to  different strengths with respect to those expressed in Eqs.~\eqref{eq:couplings_1d}-\eqref{eq:couplings_2d}. In particular, for $d=1$ we find that the expression of the external field is 
\beq
\label{eq:transverse_field}
 h_{\mathsf{a}}=2\left(m+\frac{r}{a_1}\right)\delta_{\mathsf{a},\mathsf{z}}
\eeq
whereas the spin-spin couplings following the super-exchange mechanism of virtual double occupancies now read
\beq
\label{eq:couplings_1d_wp}
J_{1,\mathsf{x}}=\frac{1-r^2}{g^2a_1},\hspace{1ex} J_{1,\mathsf{y}}=-\frac{1+r^2}{g^2a_1},\hspace{1ex} J_{1,\mathsf{z}}=-\frac{1-r^2}{g^2a_1}.
\eeq
One  can  readily see how the previous ferromagnetic Ising model in a transverse field~\eqref{eq:couplings_1d} is recovered for $r\to1$. In this limit the distinction between ferromagnetic and antiferromagnetic couplings is trivial, as one can invert the sign of the spin-spin couplings $J_{1,\mathsf{y}}\mapsto-J_{1,\mathsf{y}}$ by a unitary transformation that takes ${S}_{\mathsf{y}}(\boldsymbol{x})\mapsto(-1)^{n_1}{S}_{\mathsf{y}}(\boldsymbol{x})$. Under this transformation, the SSB ferromagnetic groundstate is transformed into a classical  N\'eel pattern of alternating spins. The discussion of the possible SSB orderings for general Wilson parameter $0<r<1$ is slightly more involved.

In the limit $r\to 0$, where one recovers the naive-fermion regularisation~\cite{gattringer_lang_2010}, the spin-spin couplings tend to $J_{1,\mathsf{x}}=-J_{1,\mathsf{y}}=-J_{1,\mathsf{z}}$,   which are  unitarily-equivalent to a  quantum Heisenberg model with antiferromagnetic couplings~\cite{Heisenberg1928,HALDANE1983464,PhysRevLett.50.1153,PhysRevLett.61.1029,Affleck1990}. The important point is that, if there is an even number of  spin-spin couplings with negative sign, these can always be inverted by a spin rotation along the remaining axis with an alternating angle. The specific transformation in this case takes  $\boldsymbol{S}(\boldsymbol{x})\mapsto({S}_{\mathsf{x}}(\boldsymbol{x}),(-1)^{n_1}{S}_{\mathsf{y}}(\boldsymbol{x}),(-1)^{n_1}{S}_{\mathsf{z}}(\boldsymbol{x}))$, such that $J_{1,\mathsf{a}}\mapsto J=1/g^2a>0$, $\forall\mathsf{a}=\{\mathsf{x,y,z}\}$, and the  spin-spin interactions clearly displays the $SU(2)$ symmetry of the Heisenberg model. Additionally, if the external field is non-zero, this maps onto a staggered transverse field under the above transformation, such that  
\beq
\label{eq:Heisenberg_1d}
H_{\rm eff}\mapsto \hat{H}_{\rm eff}=\!\!\sum_{\boldsymbol{x}\in\Lambda_1}\!\!\!\left(\!J\boldsymbol{S}(\boldsymbol{x})\cdot\boldsymbol{S}(\boldsymbol{x}+a_1\boldsymbol{e}_1)+h_\mathsf{z}\ee^{\ii\boldsymbol{k}_s\cdot\boldsymbol{x}}{S}_{\mathsf{z}}(\boldsymbol{x})\!\right),
\eeq
where we have introduced the  wavevector $\boldsymbol{k}_s=\frac{\pi}{a_1}\boldsymbol{e}_1$.
This transformation unveils  a continuous $U(1)$ symmetry with respect to rotations along the  internal $\mathsf{z}$ axis, which is not directly apparent in the original formulation~\eqref{eq:spin_model}. It is interesting to note that rewriting this transformed model in terms of Jordan-Wigner fermions maps the spin chain into a staggered-fermion regularisation~\cite{PhysRevD.11.395,PhysRevD.16.3031} of the $D=(1+1)$-dimensional Thirring model~\cite{PhysRevD.99.034504} for a single fermion flavour, provided that the four-Fermi term has a specific coupling strength. We also note that for vanishing transverse field $h_{\mathsf{z}}=0$, the Heisenberg chain was exactly solved via the Bethe ansatz~\cite{Bethe1931,hulthen1938austauschproblem}, and does not support long-range  order, as shown via the inverse scattering method~\cite{BOGOLIUBOV1986687}. 

Given the clear difference between $r=1$ and $r=0$ limits, one may expect different groundstates with distinct magnetic orders, i.e. fermion condensates, for Wilson parameters $0<r<1$. In this regime, we recall that the non-interacting phase diagram comprises regions of non-trivial topological insulators and trivial band insulators separated by topological gap-closing phase transitions~\eqref{eq:zak}. For strong interactions, the absolute values of the spin-spin couplings~\eqref{eq:couplings_1d_wp} are no longer equal, and the mapping to the antiferromagnetic Heisenberg model no longer holds. Interestingly, one can still find a $U(1)$ symmetry by combining a pair of transformations. First, rotate the spins about the internal $\mathsf{z}$ axis in an alternate fashion as $\boldsymbol{S}(\boldsymbol{x})\mapsto((-1)^{n_1}{S}_{\mathsf{x}}(\boldsymbol{x}),(-1)^{n_1}{S}_{\mathsf{y}}(\boldsymbol{x}),{S}_{\mathsf{z}}(\boldsymbol{x}))$, which effectively changes the spin-spin couplings to $J_{1,\mathsf{x}}=J_{1,\mathsf{z}}=J_{\perp}$, and $J_{1,\mathsf{y}}=J_\perp\Delta$, where 
 \beq
 \label{eq:xxz_couplings}
 J_{\perp}=\frac{r^2-1}{g^2a_1},\hspace{2ex} \Delta=\frac{1+r^2}{1-r^2}.
 \eeq
Next, apply a rotation about the internal $\mathsf{x}$ axis $\boldsymbol{S}(\boldsymbol{x})\mapsto ({S}_{\mathsf{x}}(\boldsymbol{x}),-{S}_{\mathsf{z}}(\boldsymbol{x}),{S}_{\mathsf{y}}(\boldsymbol{x}))$; the spin chain maps onto the $\mathsf{XXZ}$ model~\cite{KASTELEIJN1952104,PhysRevB.12.3908,Affleck1990}, also known as a Heisenberg-Ising model,  under an additional longitudinal field
 \beq
 \label{eq:xxz_model}
 \begin{split}
H_{\rm eff}\mapsto \hat{H}_{\rm eff}=\!J_{\perp}\!\!\!\!\sum_{\boldsymbol{x}\in\Lambda_1}\!\!\!\Big(\!{S}_\mathsf{x}(\boldsymbol{x}){S}_{\mathsf{x}}(\boldsymbol{x}+a_1\boldsymbol{e}_1)+{S}_\mathsf{y}(\boldsymbol{x}){S}_{\mathsf{y}}(\boldsymbol{x}+a_1\boldsymbol{e}_1)\\
+\Delta{S}_\mathsf{z}(\boldsymbol{x}){S}_{\mathsf{z}}(\boldsymbol{x}+a_1\boldsymbol{e}_1)+g_\mathsf{y}{S}_{\mathsf{y}}(\boldsymbol{x})\!\Big).
\end{split}
\eeq
As advanced previously, for $g_{\mathsf{y}}={h_\mathsf{z}}/{J_\perp}=0$, there is a $U(1)$ symmetry with respect to continuous rotations about the new $\mathsf{z}$ axis. In this limit, the $\mathsf{XXZ}$ model for $S=1/2$ is known to display a Berezinskii-Kosterlitz-Thouless  (BKT) phase transition~\cite{Berezinsky:1972rfj,Kosterlitz_1973} at the $SU(2)$-symmetric point $\Delta=1$~\cite{AFFLECK1985397},  separating a critical phase at $\Delta<1$ from an Ising SSB phase phase at $\Delta>1$. This model has also been exactly solved via the Bethe ansatz~\cite{PhysRev.112.309,PhysRev.116.1089}, and the inverse scattering method~\cite{BOGOLIUBOV1986687} demonstrates the different decay of spin-spin correlations in the critical and Ising phases, yielding long-range order in the latter. Following these results, we expect that such a BKT transition will not appear in the strong-coupling limit of our model~\eqref{eq:H_lattice_Wilson}, as the effective  anisotropy~\eqref{eq:xxz_couplings} always exceeds unity $\Delta>1$ for $0<r< 1$. This favours the  Ising long-range ordered phase which, after reversing the previous spin transformations, corresponds to ${\rm FM}_{\mathsf{y}}$ order, i.e. pseudo-scalar condensate in the fermion language~\eqref{eq:pseudo-sclalr_condensate}. There may be, however, other types of transition when the longitudinal field is switched on $g_{\mathsf{y}}\neq0$. Despite lacking $U(1)$ symmetry in these cases, 
the global $\mathbb{Z}_2$ symmetry~\eqref{eq:symmetry_spin_model} remains intact in the Hamiltonian~\eqref{eq:xxz_model} for any $r>0$ provided that we consider its correspondence in terms of the new rotated spin axes. Accordingly,  similar second-order quantum  phase transitions as those discussed for the quantum Ising model~\eqref{eq:couplings_1d} may still appear, albeit at  critical points that flow with  the Wilson parameter. 

\subsection{  Heisenberg-Ising compass models for $d=2$}

Before checking the validity of the above conjecture, let us  discuss the effect of non-unit Wilson parameters on the effective spin model for the $d=2$ case depicted in Fig.~\ref{fig:spin_model} {\bf (b)}. Instead of the microscopic couplings in Eq.~\eqref{eq:couplings_2d}, the super-exchange for $0<r<1$ leads to an    external field 
\beq
\label{eq:t_f_2d}
 h_{\mathsf{a}}=2\left(m+\frac{r}{a_1}+\frac{r}{a_2}\right)\delta_{\mathsf{a},\mathsf{z}},
\eeq
whereas the spin-spin couplings transform into
\beq
\begin{split}
\label{eq:couplings_2d_wp}
J_{1,\mathsf{x}}=\frac{a_2(1-r^2)}{g^2a_1},\hspace{1ex} J_{1,\mathsf{y}}=-\frac{a_2(1+r^2)}{g^2a_1},\hspace{1ex} J_{1,\mathsf{z}}=-\frac{a_2(1-r^2)}{g^2a_1},\\
J_{2,\mathsf{x}}=-\frac{a_1(1+r^2)}{g^2a_2},\hspace{1ex} J_{2,\mathsf{y}}=\frac{a_1(1-r^2)}{g^2a_2},\hspace{1ex} J_{2,\mathsf{z}}=-\frac{a_1(1-r^2)}{g^2a_2}.
\end{split}
\eeq
In the limit $r\to 1$, we recover the quantum compass model with the directional spin-spin couplings of Eq.~\eqref{eq:couplings_2d}, which supports Lorentz-breaking fermion condensates~\eqref{eq:pi_condensates} as discussed in~\cite{ziegler2020correlated,ziegler2021largen}. In  the isotropic $a_1=a_2$ and naive-fermion  $r\to 0$ limits, we recover a model that is unitarily equivalent to the  Heisenberg model on a square lattice with antiferromagnetic couplings $J_{j,\mathsf{a}}\mapsto J=1/g^2$ and staggered field $h_{\mathsf{z}}\mapsto h_{\mathsf{z}}(-1)^{n_1+n_2}$. This requires a slightly more involved  transformation in comparison  to $d=1$~\eqref{eq:Heisenberg_1d}. For odd rows, the spins must be transformed as
$
\boldsymbol{S}(\boldsymbol{x})\mapsto({S}_{\mathsf{x}}(\boldsymbol{x}),(-1)^{n_1+1}{S}_{\mathsf{y}}(\boldsymbol{x}),(-1)^{n_1+1}{S}_{\mathsf{z}}(\boldsymbol{x})),
$
$\forall \boldsymbol{x}=(n_1a_1,(2n_2-1)a_2)$, whereas for even rows they transform as  $
\boldsymbol{S}(\boldsymbol{x})\mapsto((-1)^{n_1}{S}_{\mathsf{x}}(\boldsymbol{x}),{S}_{\mathsf{y}}(\boldsymbol{x}),(-1)^{n_1}{S}_{\mathsf{z}}(\boldsymbol{x}))
$, $\forall \boldsymbol{x}=(n_1a_1,(2n_2)a_2)$, where $(n_1,n_2)\in\mathbb{Z}_{N_1}\times\mathbb{Z}_{N_2}$. This resulting Hamiltonian is
\beq
H_{\rm eff}\mapsto \hat{H}_{\rm eff}=\!\!\sum_{\boldsymbol{x}\in\Lambda_1}\!\sum_{j={1,2}}\!\!\left(\!J\boldsymbol{S}(\boldsymbol{x})\cdot\boldsymbol{S}(\boldsymbol{x}+a_j\boldsymbol{e}_j)+h_\mathsf{z}\ee^{\ii\boldsymbol{k}_s\cdot\boldsymbol{x}}{S}_{\mathsf{z}}(\boldsymbol{x})\!\right),
\eeq
where  the corresponding staggering wave-vector now reads $\boldsymbol{k}_s=\frac{\pi}{a_1}\boldsymbol{e}_1+\frac{\pi}{a_2}\boldsymbol{e}_2$. For vanishing transverse field  $h_{\mathsf{z}}=0$, one finds that the strong-coupling limit of the naive-fermion $r= 0$ isotropic limit $a_1=a_2$ corresponds exactly to the 
 two-dimensional antiferromagnetic Heisenberg model on a square lattice. This model is no longer solvable via Bethe ansatz~\cite{Bethe1931}, and has been the subject of intense research in the past~\cite{RevModPhys.63.1}. In contrast to $d=1$, all analytic and numerical evidence supports a groundstate displaying long-range anti-ferromagnetic order in this case.

Let us now discuss the  case of non-zero Wilson parameter $0<r< 1$, where Eq.~\eqref{eq:couplings_2d_wp} leads to directional Heisenberg-Ising anisotropies $|J_{1,\mathsf{y}}|>|J_{1,\mathsf{x}}|=|J_{1,\mathsf{z}}|$ and  $|J_{2,\mathsf{x}}|>|J_{2,\mathsf{y}}|=|J_{2,\mathsf{z}}|$. In the limit $a_1/a_2\to 0$ ($a_1/a_2\to \infty$), we again have a collection of uncoupled spin chains along the rows (columns), as discussed for the quantum compass model~\eqref{eq:couplings_2d}, with the difference that each of these rows (columns) is no longer described by a quantum Ising model~\eqref{eq:couplings_2d} but, instead, by an $\mathsf{XYZ}$ chain~\cite{PhysRevLett.26.834,BAXTER1972323}. As remarked around Eq.~\eqref{eq:xxz_model}, we find an even number of negative spin-spin couplings in these rows (columns), such that a  specific spin rotation maps this model to an antiferromagnetic $\mathsf{XXZ}$ model where the Ising anisotropy is always larger than unity. In analogy to the $d=1$ case discussed above, we also expect to find either a ferromagnetic  FM$_{\mathsf{x}}$ ( FM$_{\mathsf{y}}$) ordering for each of the columns (rows), or, otherwise, a symmetric paramagnet (PM) when  the transverse field is larger than  the leading spin-spin coupling.  Also following this analogy, as we depart from the limits of large anisotropies $a_1/a_2\to 0$ ($a_1/a_2\to \infty$),  the   additional spin-spin couplings will lock these columns (rows) to the same ferromagnetic order, and we expect that the critical points   of the purely 90$^{\rm o}$ compass model $|h_{\mathsf{z}}|<|J_{2,\mathsf{x}}|/\zeta$ and $|J_{2,\mathsf{x}}|>|J_{1,\mathsf{y}}|$  for the FM$_{\mathsf{x}}$-PM transition, and $|h_{\mathsf{z}}|<|J_{1,\mathsf{y}}|/\zeta$  and $|J_{1,\mathsf{y}}|>|J_{2,\mathsf{x}}|$ for FM$_{\mathsf{y}}$-PM transition will change, such that $\zeta$ flows with the Wilson parameter. 
Since these  Heisenberg-Ising compass models   are no longer solvable, this  can only be determined numerically, or using certain approximations that are discussed  below. We will start by deriving a path-integral representation of these effective spin models that will connect to variants of the constrained QFTs discussed previously.

\section{\bf $\mathbb{Z}_2$ non-linear sigma models}

\begin{figure}[t]
	\centering
	\includegraphics[width=0.5\textwidth]{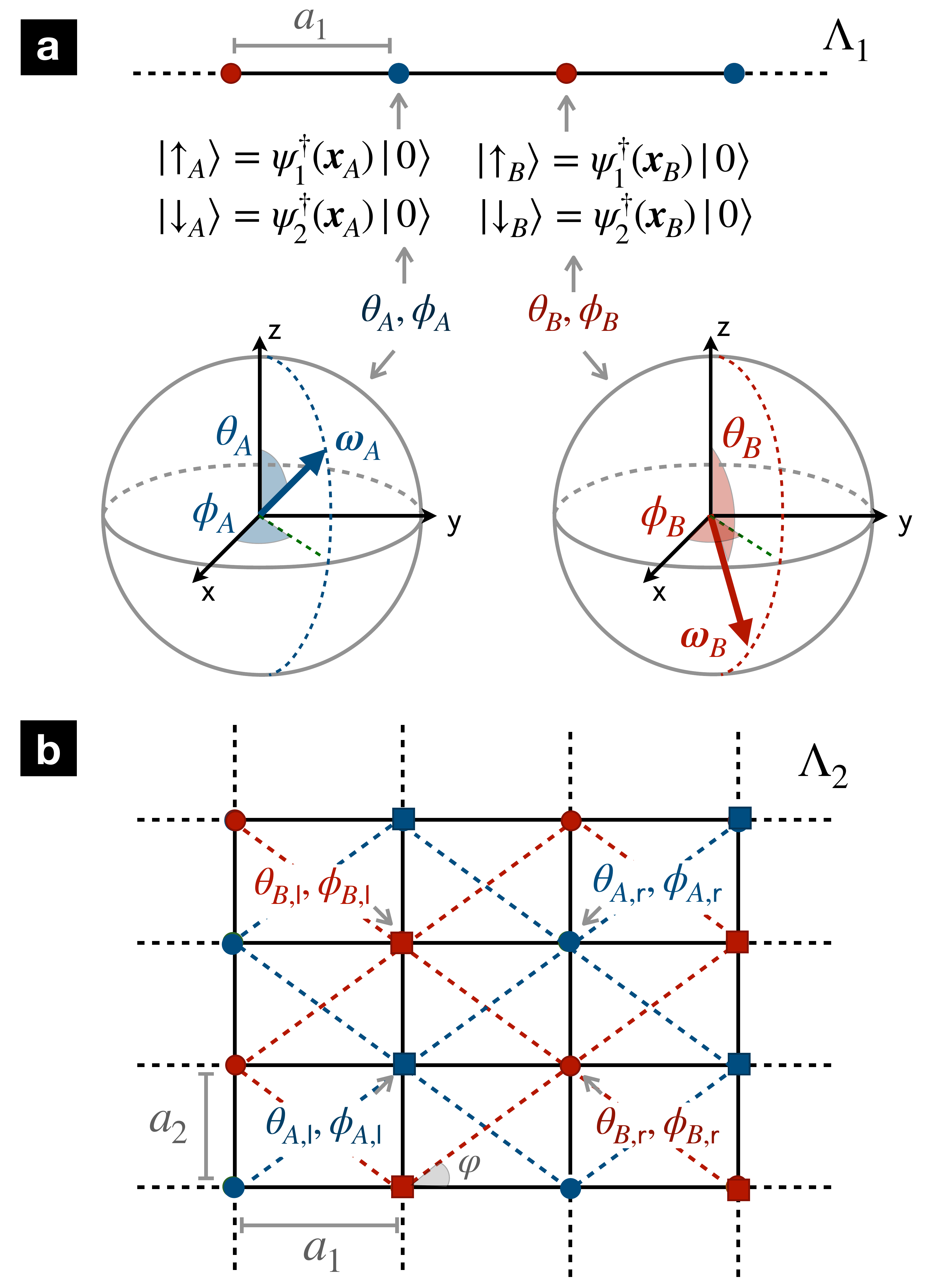}
	\caption{  {\bf Spin coherent states for bipartite lattices:} {\bf (a)} The one-dimensional chain $\Lambda_1$ of lattice spacing $a_1$ contains a 2-site unit cell with $s=A$ (odd) and $s=B$(even) sites  represented by blue and red circles. The low-energy properties of  the half-filled chain in the strong-coupling limit  can be spanned   by the $\ket{\uparrow_s},\ket{\downarrow_s}$ states, which correspond to the north and south poles of the respective $S_2$ spheres. A generic spin coherent state can be described by the unit vector $\boldsymbol{\omega}_s$ with angles $\theta_s,\phi_s$. {\bf (b)} For a rectangular lattice $\Lambda_2$ with  spacings $a_1,a_2$, the two sub-lattices $A,B$ are represented with blue and red symbols. The spin-coherent state basis is now composed of a 4-site unit cell where, in addition to the sub-lattice label $s=A,B$, we consider the left- and right- corners $c={\mathsf{l},\mathsf{r}}$, leading to $\theta_{s,c},\phi_{s,c}$ .}
	\label{fig:lattices}
\end{figure}

In this section, we derive a path-integral representation for the partition function $Z={\rm Tr}\{\ee^{-\beta H_{\rm eff}}\}$ of the strong-coupling Heisenberg-Ising model~\eqref{eq:spin_model}, which will allow us to understand how constraints in QFTs such as  Eq.~\eqref{eq:constraint} below can appear in our fermionic model~\eqref{eq:H_lattice_Wilson}. This requires the use of {\em spin coherent states}~\cite{Radcliffe_1971}, which can be defined by performing a  $SU(2)$ rotation on a fiducial state, which fulfils $\boldsymbol{S}^2(\boldsymbol{x})
\ket{S,+S}_{\boldsymbol{x}}=S(S+1)\ket{S,+S}_{\boldsymbol{x}}$ and ${S}_{\mathsf{z}}(\boldsymbol{x})
\ket{S,+S}_{\boldsymbol{x}}=S\ket{S,+S}_{\boldsymbol{x}}$.  The coherent-state basis, depicted in Fig.~\ref{fig:lattices}, can thus be defined by the action of the following operator on  the tensor product of fiducial states for each lattice site
\beq
\ket{\{\boldsymbol{\omega}(\tau,\boldsymbol{x})\}}=\ee^{\sum\limits_{\boldsymbol{x}\in\Lambda_{d}}\hspace{-1.25ex}\ii\theta\!(\tau,\boldsymbol{x})\left(\sin\phi\!(\tau,\boldsymbol{x}) S_{\mathsf{x}}(\boldsymbol{x})-\cos\phi\!(\tau,\boldsymbol{x}) S_{\mathsf{y}}(\boldsymbol{x})\right)}\bigotimes\limits_{\boldsymbol{x}\in\Lambda_d}\ket{S,S}_{\boldsymbol{x}}.
\eeq
Here, we have introduced polar $\theta({x})$ and azimuthal $\phi({x})$ angles, which define a unit vector per spin pointing along  the radial outward direction of  a unit 2-sphere $S_2$. Therefore,  $|\boldsymbol{\omega}(\tau,\boldsymbol{x})|^2=1$ parametrises all possible spin directions
\beq
\label{eq:generalised_coordinates}
\boldsymbol{\omega}({x})=\sin\theta\!({x})\cos\phi\!({x})\boldsymbol{e}_{\mathsf{x}}+\sin\theta\!({x})\sin\phi\!({x})\boldsymbol{e}_{\mathsf{y}}+\cos\theta\!({x})\boldsymbol{e}_{\mathsf{z}}.
\eeq
Note that $x=(\tau,\boldsymbol{x})$ now represents the Wick-rotated spacetime points, where imaginary time $\tau$ extent is related to inverse temperature via $\tau\in[0,\beta]$~\cite{fradkin_2013}.
One readily checks that in this basis $\langle\boldsymbol{S}(\boldsymbol{x})\rangle=S\boldsymbol{\omega}(\boldsymbol{x})$, such that the components of this unit vector field contain information about the fermionic $\sigma$ and $\pi$ condensates mentioned above
\beq
\label{eq:sigma_equivalence}
\sigma({x})={\omega}_{\mathsf{z}}({x}),\hspace{1ex} 
\boldsymbol{\pi}({x})={\omega}_{\mathsf{x}}({x})\boldsymbol{e}_{\mathsf{x}}+{\omega}_{\mathsf{y}}({x})\boldsymbol{e}_{\mathsf{y}}.
\eeq
One can now rewrite the partition function as a path integral 
 \beq
 \label{eq:measure}
 Z=\int[D\Omega]\ee^{S_{E}},\hspace{1ex}[D\Omega]=\prod\limits_{\boldsymbol{x}\in\Lambda_d}\!{\rm d}^3\Omega\frac{(2S+1)}{4\pi}\delta(\boldsymbol{\Omega}^2(\tau,\boldsymbol{x})-1),
 \eeq
 where the vector field $\Omega(\tau,\boldsymbol{x})$ is constrained to lie on the $S_2$ sphere 
 \beq
 \label{eq:constraint}
 \boldsymbol{\Omega}^2(\tau,\boldsymbol{x})=1,\hspace{2ex}\forall\boldsymbol{x}\in\Lambda_d,
  \eeq
  through the specific form of the  integral measure. As discussed in~\cite{RevModPhys.63.1,fradkin_2013, PhysRevLett.61.1029}, the Euclidean action contains a geometric contribution proportional to the sum of the  Berry phases of each spin history as it   moves along  the corresponding trajectory $\Gamma_{\boldsymbol{x}}:\tau\to\boldsymbol{\Omega}(\tau,{\boldsymbol{x}})$ on its respective  sphere, these trajectories being closed due to the periodic boundary conditions along the $\tau$ direction $\boldsymbol{\Omega}(0,{\boldsymbol{x}})=\boldsymbol{\Omega}(\beta,{\boldsymbol{x}})$. Altogether, the action is expressed as 
 \begin{widetext}
 \beq
  \label{action_eff_spin_model}
 S_E=\sum_{\boldsymbol{x}\in\Lambda_d}\left(-\ii\!\oint_{\Gamma_{\boldsymbol{x}}}\!{\rm d}\boldsymbol{\Omega}\cdot\boldsymbol{A}(\boldsymbol{\Omega}(\tau,\boldsymbol{x}))+\int_0^\beta\!\!\!{\rm d}\tau\sum_{\mathsf{a}}\left(\sum_jJ_{j,\mathsf{a}}S^2\Omega_\mathsf{a}(\tau,\boldsymbol{x})\Omega_\mathsf{a}(\tau,\boldsymbol{x}+a_j\boldsymbol{e}_j)+h_{\mathsf{a}}S\Omega_\mathsf{a}(\tau,\boldsymbol{x})\right)\right),
 \eeq
  \end{widetext}
 where the first contribution corresponds to the aforementioned Berry phase~\cite{berry_phase_review}, and is known as a Wess-Zumino term. For each spin, this term can be understood as an effective Aharonov-Bohm phase  gained by a  unit test charge $q_e=1$ moving on the sphere, and subjected to the magnetic field of a monopole of charge $q_m=4\pi S$ located at its centre~\cite{nakahara_2017}. Using Stokes' theorem, this phase can be rewritten as the magnetic flux across the  spherical cap enclosed by each spin trajectory containing the north pole  of $S_2$, where the fiducial state $\ket{S,+S}_{\boldsymbol{x}}$ points to. Hence,   the effective vector potential and magnetic field are those generated by the magnetic monopole
 \beq
 \label{eq:magnatic_monopole}
 \begin{split}
 \boldsymbol{A}(\boldsymbol{\Omega}(\tau,\boldsymbol{x}))&=\frac{S}{|\boldsymbol{\Omega}(\tau,\boldsymbol{x})|}\frac{\boldsymbol{e}_{\mathsf{z}}\times\boldsymbol{\omega}(\tau,\boldsymbol{x})}{1+{\boldsymbol{e}_{\mathsf{z}}\cdot\boldsymbol{\omega}(\tau,\boldsymbol{x})}},\\
 \boldsymbol{B}(\boldsymbol{\Omega}(\tau,\boldsymbol{x}))&=\boldsymbol{\nabla}_{\boldsymbol{\Omega}}\times \boldsymbol{A}=\frac{S \boldsymbol{\omega}(\tau,\boldsymbol{x}) }{|\boldsymbol{\Omega}(\tau,\boldsymbol{x})|^2}.
 \end{split}
\eeq 
 The second term in ~\eqref{action_eff_spin_model} represents the additional coupling of neighbouring spins due to the spin-spin couplings, as well as  their precession under the transverse  field. 
 
 Let us briefly discuss the $r\to 0$ naive-fermion limit, and its relation to an $O(3)$ non-linear sigma model~\cite{PhysRevLett.61.1029}. In this limit,  the spin-spin couplings along the different internal directions are all equal $|J_{j,\mathsf{a}}|=J_j$, and there is a continuous $SU(2)$ symmetry. In light of the field constraint in the integral measure~\eqref{eq:measure}, and up to an irrelevant constant term,  the nearest-neighbour couplings can be rewritten as follows  $-J_ja_j\frac{1}{2a_j}(\boldsymbol{\Omega}(\tau,\boldsymbol{x}+a_j\boldsymbol{e}_j)-\boldsymbol{\Omega}(\tau,\boldsymbol{x}))^2\to \frac{J_ja_j}{2}\partial^j\boldsymbol{\Omega}\cdot\partial_j\boldsymbol{\Omega}$, which clearly resembles the spatial derivative-terms in Eq.~\eqref{eq:H_0}, and can be understood as the energy contribution due to the strain caused by a field deformation. The kinetic part, which depends on the canonical momenta of the scalar vector field in Eq.~\eqref{eq:H_0}, would appear in the form of Euclidean time derivatives in the corresponding action, and is not readily apparent in  equation~\eqref{action_eff_spin_model}. However, a specific parametrisation of the spin trajectory shows that the Berry phase indeed contains these time derivatives $\int_{0}^\beta\!{\rm d}\tau\boldsymbol{A}(\boldsymbol{\Omega})\cdot\partial_\tau\boldsymbol{\Omega}$, albeit still being different from the kinetic terms of the $O(3)$ non-linear sigma model.  In a seminal work~\cite{PhysRevLett.61.1029}, F.D.M. Haldane showed that for anti-ferromagnetic Heisenberg couplings $J_{j}=J>0$, an expansion of  $\boldsymbol{\Omega}(\tau,\boldsymbol{x})$ about the saddle point of the Euclidean action, corresponding to the alternating antiferromagnetic configuration of the classical limit, exactly yields the kinetic term of the $O(3)$ non-linear sigma model. Moreover, in $d=1$, the Berry phase also contributes with a topological theta term $\theta=2\pi S$ which, depending on the half-integer or integer value of the spin $S$, makes the $O(3)$ non-linear sigma model massless or massive~\cite{PhysRevB.48.3844}. 
 
 Let us now explore how this situation changes for our current spin models. Since the effective spin models do not have the continuous $SU(2)$ symmetry of the Heisenberg limit for generic $0<r<1$, we first need to understand the nature of the saddle point of Eq.~\eqref{action_eff_spin_model} controlling the large-$S$ limit, which will eventually lead to a different type of constrained non-linear sigma model. By inspection of the action~\eqref{action_eff_spin_model} and the magnetic monopole fields~\eqref{eq:magnatic_monopole}, one can readily see that the ${\rm d}\boldsymbol{\Omega}\cdot\boldsymbol{A}(\boldsymbol{\Omega}(\tau,\boldsymbol{x}))$ term can only depend on the equatorial components of the scalar field via  $\sum_{\boldsymbol{x}\in\Lambda_d}(\Omega_{\mathsf{x}}\partial_\tau\Omega_{\mathsf{y}}-\Omega_{\mathsf{y}}\partial_\tau\Omega_{\mathsf{x}})$. This term remains invariant under the $\mathbb{Z}_2$ symmetry~\eqref{eq:symmetry_spin_model} which in this context reads
 \beq
\boldsymbol{\Omega}(\tau,\boldsymbol{x})\mapsto(-{\Omega}_{\mathsf{x}}(\tau,-\boldsymbol{x}),-{\Omega}_{\mathsf{y}}(\tau,-\boldsymbol{x}),{\Omega}_{\mathsf{z}}(\tau,-\boldsymbol{x})).
\eeq
Accordingly,  the action~\eqref{action_eff_spin_model} with the constraint ~\eqref{eq:measure} (equivalent to the non-linear sigma constraint~\eqref{eq:constraint} via Eq.~\eqref{eq:sigma_equivalence}) describes a {\it $\mathbb{Z}_2$ non-linear sigma model} that arises naturally in the strong-coupling limit of the original model~\eqref{eq:H_lattice_Wilson}, even for a single fermion  flavour $N=1$.

\section{\bf Large-$S$ limit and saddle-point equations}

Let us recall that the spin operators in Eq.~\eqref{eq:spin_operators} correspond to $S=1/2$. In this section, nonetheless, we will assume that $S$ is a free parameter, and explore the $S\to\infty$ limit, which can be understood as a mean-field approximation of the effective spin chain. According to our previous discussion of the Berry phase,  one can see that  time-dependent spin histories $\partial_\tau\boldsymbol{\Omega}\neq \boldsymbol{0}$ will get suppressed in this limit, due to the averaging of the rapid oscillations  associated to the pure-imaginary Wess-Zumino term. Accordingly, the large-$S$ limit will be controlled by static fields $\boldsymbol{\Omega}(\tau,\boldsymbol{x})=\boldsymbol{\Omega}(\boldsymbol{x})$. 
This is precisely analogous to the large-$N$ limit in interacting fermion theories; e.g. in the $d=2$ Thirring model the induced Chern-Simons term resulting from the leading quantum correction~\cite{Gomes:1990ed} plays no role in determining the ground state in the large-$N$ limit.
Moreover, the Euclidean action can be rewritten as $S_E=\frac{1}{\hbar_{\rm eff}}s_E$ where ${\hbar_{\rm eff}}\propto 1/S$ plays the role of an effective Planck constant. In the absence of dynamics and kinetic terms, the  Euclidean action per spin $s_E=\beta V_{\rm eff}$ can  be expressed in terms of the following effective potential
\beq
\label{eq:eff_potential}
V_{\rm eff}(\{\boldsymbol{\Omega}\})=\sum_{\boldsymbol{x}\in\Lambda_d}\sum_{\mathsf{a}}\left(\sum_j\tilde{J}_{j,\mathsf{a}}\Omega_\mathsf{a}(\boldsymbol{x})\Omega_\mathsf{a}(\boldsymbol{x}+a_j\boldsymbol{e}_j)+h_{\mathsf{a}}\Omega_\mathsf{a}(\boldsymbol{x})\!\!\right),
\eeq
 where we note that, in analogy to the large-$N$ limit~\cite{coleman_1985} of the four-Fermi term~\eqref{eq:int_H_D}, the spin-spin couplings~\eqref{eq:couplings_1d_wp}-\eqref{eq:couplings_2d_wp}  must be rescaled so as to give finite contributions for $S\to\infty$
\beq
J_{j,\mathsf{a}}=\frac{\tilde{J}_{j,\mathsf{a}}}{S},
\eeq
where $\tilde{J}_{j,\mathsf{a}}$ are finite and non-zero coupling strengths.

Accordingly, the large-$S$ limit is controlled by the saddle point  of the potential~\eqref{eq:eff_potential} given below, in which quantum fluctuations are suppressed ${\hbar_{\rm eff}}\to 0$, bearing in mind that the fields are subjected to an extensive number of constraints, i.e. one per spatial coordinate, as they must lie on their corresponding $S_2$ spheres~\eqref{eq:constraint}.  At this level, one can either introduce a Lagrange multiplier to deal with the non-linear constraint, or work directly with the generalised 'coordinates' $\{\boldsymbol{\Omega}(\boldsymbol{x})\}\mapsto\{\theta(\boldsymbol{x}),\phi(\boldsymbol{x})\}$  satisfying the constraints~\eqref{eq:generalised_coordinates}. In this second approach, the saddle-point equations are given by a set of $2\prod_jN_j$ non-linear equations $\forall\boldsymbol{x}\in\Lambda_d$, namely
\beq
\label{eq:saddle_point_conditions}
\left.\frac{\partial V_{\rm eff}(\{\theta\!(\boldsymbol{x}),\phi\!(\boldsymbol{x})\})}{\partial{\theta\!(\boldsymbol{x})}}\right|_{\theta^\star,\phi^\star}\!\!=\left.\frac{\partial V_{\rm eff}(\{\theta\!(\boldsymbol{x}),\phi\!(\boldsymbol{x})\})}{\partial{\phi\!(\boldsymbol{x})}}\right|_{\theta^\star,\phi^\star}\!\!=0.
\eeq

\subsection{Large-$S$ Ising  magnetism for $d=1$}
\label{sec:largeS_d1}

Let us start by discussing the solutions of these saddle-point equations for $d=1$.
Following the discussion in Sec.~\ref{sec:Heis_d_1}, where we argued that the spin couplings~\eqref{eq:couplings_1d_wp} for $0<r< 1$ can be mapped onto an antiferromagnetic Heisenberg-Ising chain~\eqref{eq:xxz_model},  we can simplify the set of non-linear equations~\eqref{eq:saddle_point_conditions} by restricting to translationally-invariant configurations within a $2$-site unit cell $\{\theta(\boldsymbol{x}),\phi(\boldsymbol{x})\}\mapsto\{\theta_{s},\phi_{s}\}$, which may capture a possible alternating order, where  $s\in\{A,B\}$ stands for the odd/even sites of the chain (see Fig.~\ref{fig:lattices}{\bf (a)}). Under this simplification, the effective potential reads

\begin{widetext}
\beq
\label{eq:eff_potential}
\frac{V_{\rm eff}(\theta_A,\theta_B,\phi_A,\phi_B)}{N_1}=\sin\theta_{A}\sin\theta_{B}\left(\tilde{J}_{1,\mathsf{x}}\cos\phi_{A}\cos\phi_{B}+\tilde{J}_{1,\mathsf{y}}\sin\phi_{A}\sin\phi_{B}\right)+\tilde{J}_{1,\mathsf{z}}\cos\theta_{A}\cos\theta_{B}+\half h_{\mathsf{z}}(\cos\theta_{A}+\cos\theta_{B}),
\eeq
\end{widetext}
which leads to a system of four non-linear equations via ~\eqref{eq:saddle_point_conditions}. To gain some knowledge about the groundstate, one can numerically search for a global minimum of this potential using a coarse-grained discretization of the angles, and performing a grid search restricting the search space to account for the $\mathbb{Z}_2$ symmetry. Once we have a rough estimate of the minimum, for further accuracy  we directly minimise the effective potential as a   non-linear constrained problem using an interior-point algorithm of non-linear programming~\cite{NoceWrig06}, choosing as initial points the outcomes of the global coarser minimisation. In practice, we have also initialised the minimisation in all  the different combinations of cardinal states of the two $S_2$ spheres  associated to the 2-site unit cell, and checked for consistency, ensuring that the solution found with the non-linear programming algorithm that starts with the grid-search  minimum yields the minimum  potential among the different starting points. This analysis yields saddle-point solutions $\{\theta_A^\star,\theta_B^\star, \phi_A^\star,\phi_B^\star\}$ for different values of the Wilson parameter $r\in\{0.5,0.6,0.7,0.8,0.9,1\}$.

 \begin{figure}[t]
	\centering
	\includegraphics[width=0.4\textwidth]{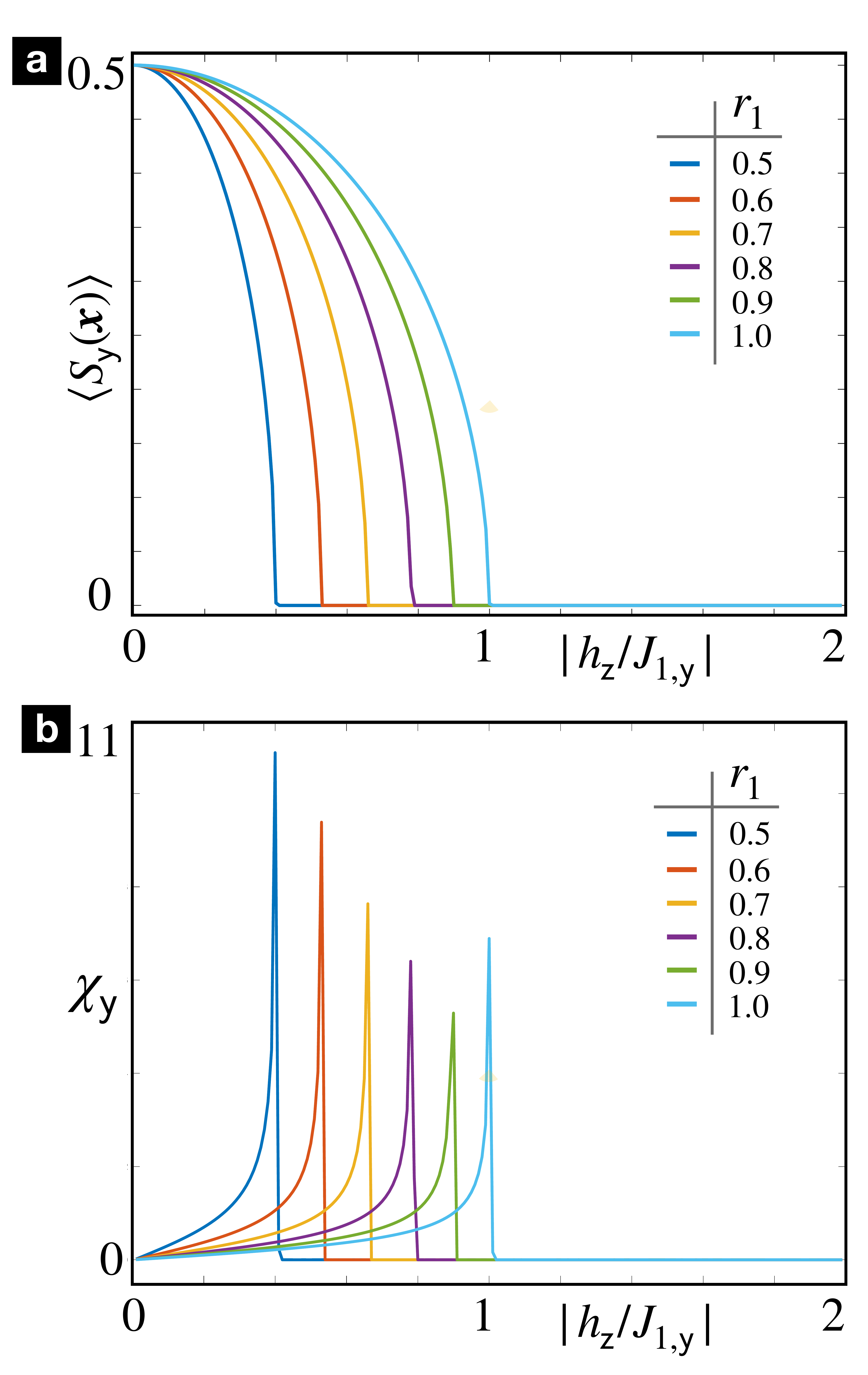}
	\caption{{\bf Large-$S$ Ising magnetism  in the Heisenberg-Ising chain for $d=1$:} {\bf (a)} Ferromagnetic order parameter $\langle S_{\mathsf{y}}(\boldsymbol{x})\rangle$ as a function of the relative transverse field, considering various values of the Wilson parameter. For each value of $r$, the region with a non-zero magnetisation corresponds to the long-range ordered FM$_{\mathsf{y}}$. {\bf (b)} Chiral magnetic susceptibility $\chi_{\mathsf{y}}=\partial\langle S_{\mathsf{y}}(\boldsymbol{x})\rangle /\partial h_{\mathsf{z}}$, which peaks at the critical points of SSB. 
	}
	\label{fig:d_1_ssb}
\end{figure}

In Fig.~\ref{fig:d_1_ssb} {\bf (a)}, we use these numerical large-$S$ solutions to  represent the corresponding SSB order parameter, which corresponds to the pseudo-scalar condensate  $\langle S_{\mathsf{y}}(\boldsymbol{x})\rangle \propto\Pi_5 $ in Eq.~\eqref{eq:pseudo-sclalr_condensate}. This figure clearly depicts a SSB region hosting an Ising ferromagnet FM$_{\mathsf{y}}$, corresponding to the parity-breaking Aoki phase with a non-zero pseudo-scalar condensate. This phase is separated from the one invariant under parity, namely a disordered paramagnet PM in the language of the spin model, via a critical point where the order parameter behaves non-analytically. To find the accurate location of this point, we represent in Fig.~\ref{fig:d_1_ssb} {\bf (b)} the corresponding susceptibilities $\chi_{\mathsf{y}}=\partial\langle S_{\mathsf{y}}(\boldsymbol{x})\rangle /\partial h_{\mathsf{z}}$, which clearly peak at the corresponding points. These figures  show, as qualitatively argued in the previous section,that the critical point $h_{\mathsf{z}}/J_{1,\mathsf{y}}\big|_{\rm c}$ flows with the value of the Wilson parameter $r$.  For $r=1$, the critical point  obtained from the numerical minimisation is $|h_{\mathsf{z}}/J_{1,\mathsf{z}}|_{\rm c}\approx 1$. This coincides with  the analytical  solution of the saddle-point equations which, in this limit, can be found exactly
\beq
\phi_A^\star=\phi_B^\star\in\left\{\!\frac{\pi}{2},\frac{3\pi}{2}\!\right\},\hspace{1ex}\theta_A^\star=\theta_B^\star=\pi-\arccos\left(\frac{h_{\mathsf{z}}}{|J_{1,\mathsf{y}}|}\right).
\eeq
We thus find that, for $|h_{\mathsf{z}}/J_{1,\mathsf{y}}|<|h_{\mathsf{z}}/J_{1,\mathsf{y}}|_{\rm c}=1$,  the spins align according to 
\beq
\label{eq:solution_r_1}
\langle\boldsymbol{S}(\boldsymbol{x})\rangle=\pm S\sqrt{1-\left(\frac{h_{\mathsf{z}}}{J_{1,\mathsf{y}}}\right)^{\!\!\!2}}\boldsymbol{e}_{\mathsf{y}}-S\frac{h_{\mathsf{z}}}{|J_{1,\mathsf{y}}|}\boldsymbol{e}_{\mathsf{z}},
\eeq
where the two possible signs $\pm$ account for the two-fold degeneracy associated to the $\mathbb{Z}_2$ parity SSB.

\begin{figure}[t]
	\centering
	\includegraphics[width=0.5\textwidth]{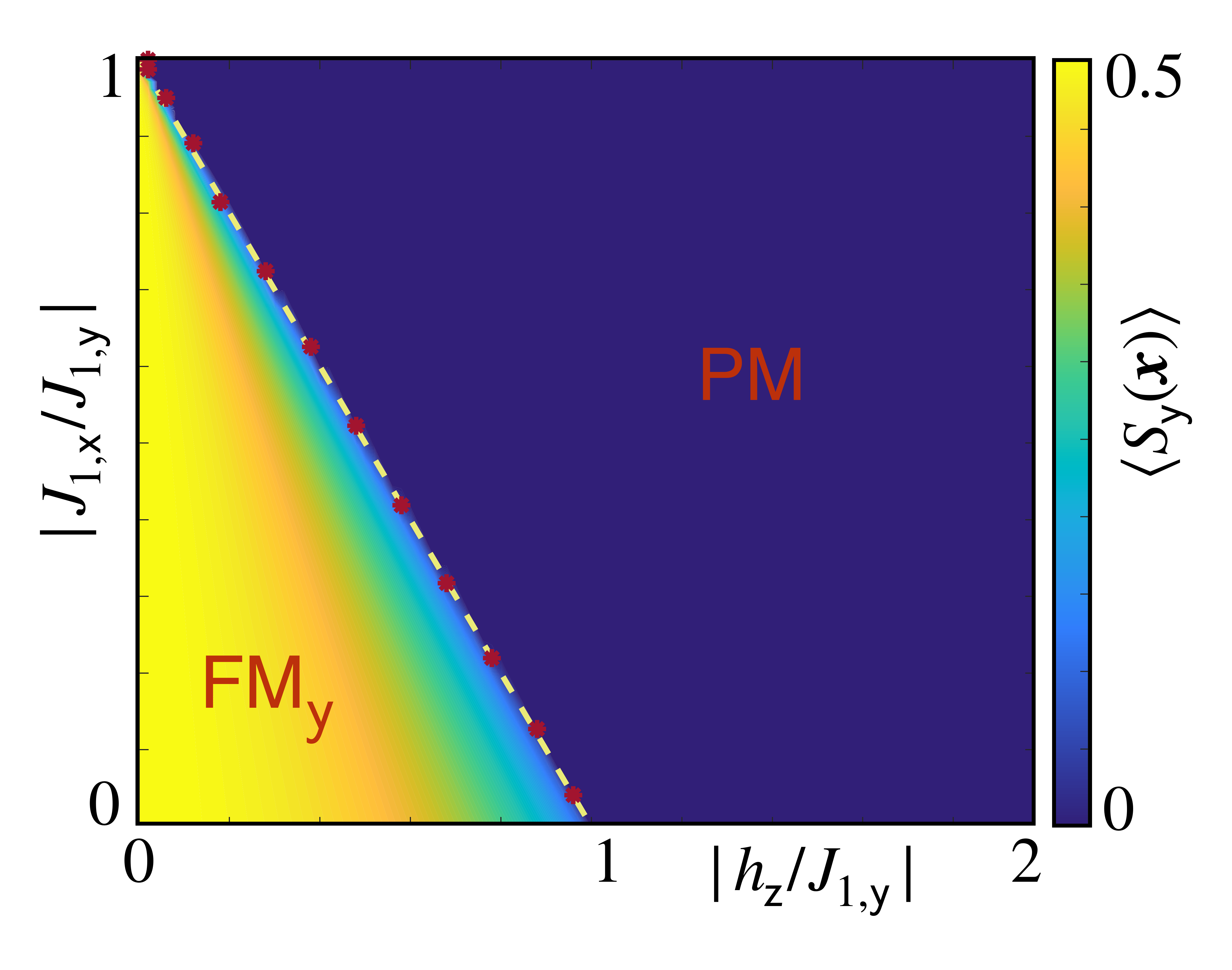}
	\caption{{\bf Large-$S$ phase diagram for  $d=1$:}  Magnetisation contour plot, including the red stars that stand for the critical points obtained numerically in Fig.~\ref{fig:d_1_ssb}{\bf (b)}, as well as the white dashed line for the predictions in Eq.~\eqref{eq:mf_prediction}, which correspond to a straight line with negative unit slope $\left|{h_{\mathsf{z}}}/{J_{1,\mathsf{y}}}\right|_{\rm c}=1-|J_{1,\mathsf{x}}/J_{1,\mathsf{y}}|$.   }
	\label{fig:phase_diagram_d_1}
\end{figure}

As one now varies $0<r<1$, it is simple to understand the flow of the critical point by performing a self-consistent mean-field decoupling of the effective Hamiltonian~\eqref{eq:spin_model}. In light of the groundstate expectation values ~\eqref{eq:solution_r_1}, a mean-field decoupling of the additional terms of the Heisenberg-Ising chain~\eqref{eq:xxz_model} that arise when $r \neq 1$
\beq
\label{eq:mf_decoupling}
\sum\limits_{\mathsf{a}=\mathsf{x,z}}{J}_{1,\mathsf{a}}S_{\mathsf{a}}(\boldsymbol{x})S_{\mathsf{a}}(\boldsymbol{x}+a_1\boldsymbol{e}_1)\mapsto \sum_{\mathsf{a}=\mathsf{x,z}}2{J}_{1,\mathsf{a}}S_{\mathsf{a}}(\boldsymbol{x})\langle S_{\mathsf{a}}(\boldsymbol{x})\rangle,
\eeq
would only contribute with  terms along the internal $\mathsf{z}$ direction, effectively shifting the transverse field to $h_{\mathsf{z}}\mapsto \tilde{h}_{\mathsf{z}}= h_{\mathsf{z}}+2{J}_{1,\mathsf{z}}\langle S_{\mathsf{z}}(\boldsymbol{x})\rangle$.   The saddle-point solution~\eqref{eq:solution_r_1} now yields  a self-consistent equation, which can be readily solved for the couplings in Eq.~\eqref{eq:couplings_1d_wp}:
\beq
\label{eq:mf_prediction}
\left|\frac{h_{\mathsf{z}}}{J_{1,\mathsf{y}}}\right|_{\rm c}=1-\frac{1-r^2}{1+r^2}.
\eeq
In order to test the validity of this prediction, we present a contour plot of the Ising order parameter in Fig.~\ref{fig:d_1_ssb} {\bf (c)} as a function of the relative coupling strengths $|{J}_{1,\mathsf{x}}/{J}_{1,\mathsf{y}}|=|{J}_{1,\mathsf{z}}/{J}_{1,\mathsf{y}}|=(1-r^2)/(1+r^2)$ and $|{h}_{\mathsf{z}}/{J}_{1,\mathsf{y}}|$. We  also plot the critical points extracted from the numerical maxima of the susceptibility in 
Fig.~\ref{fig:phase_diagram_d_1},
depicted with  red stars, and their analytical prediction~\eqref{eq:mf_prediction}, represented with a white dashed line. The agreement is very good, and identifies the main source for the flow of the critical coupling with the Wilson parameter. In the $d=1$ case, the critical  point flows towards smaller values $|h_{\mathsf{z}}/J_{1,\mathsf{y}}|_{\rm c}<1$ due to the effective renormalisation of the transverse field $h_{\mathsf{z}}\mapsto \tilde{h}_{\mathsf{z}}$, which  increases $\tilde{h}_{\mathsf{z}}>h_{\mathsf{z}}$ due to the non-vanishing ferromagnetic couplings $J_{1,\mathsf{z}}<0$ along the internal $\mathsf{z}$ axis,  and the specific alignment of the spins in Eq.~\eqref{eq:solution_r_1}. Coming back to the context of four-Fermi-Wilson lattice field theories~\eqref{eq:H_lattice_Wilson}, we see that the Aoki phase extends for arbitrary values of the Wilson parameter $0<r<1$, provided that one goes sufficiently close to the so-called central Wilson branch $m=-1/a_1$~\cite{10.1093/ptep/ptaa003,PhysRevD.102.034516}. For strictly vanishing $r=0$, where we recover the naive-fermion discretization, the Aoki phase is no longer present. Right at the central branch, where $h_{\mathsf{z}}=0$, the strong-coupling groundstate lies exactly at the critical line, and would correspond to the gapless  groundstate of the antiferromagnetic Heisenberg chain.

Before moving towards the $d=2$ case, let us note that the large-$S$ approximation is not expected to provide an accurate estimate of the exact critical point but, at least, it  captures qualitatively the main sources for the flow of the critical point. In fact, for  the quantum Ising model at $r=1$, where large-$S$ predicts a  critical point  $|h_{\mathsf{z}}/J_{1,\mathsf{z}}|=1$ in Fig.~\ref{fig:d_1_ssb}{\bf (a)-{\bf(b)}},  the exact solution gives instead instead  $|h_{\mathsf{z}}/J_{1,\mathsf{z}}|=1/2$~\cite{PFEUTY197079}. In  section~\ref{sec:tn}, we will present more accurate predictions of the critical points using the quasi-exact DMRG algorithm based on matrix product states.

\subsection{ Large-$S$ compass magnetism for $d=2$}
\label{sec:2d_largeS}

Let us now discuss the $d=2$ case, where  the $A$ and $B$ sub-lattices correspond to two interpenetrating square lattices with  lattice spacing $(a_1^2+a_2^2)^{1/2}$, rotated with respect to the original rectangular lattice by angles $\varphi,\pi-\varphi$, where  $\varphi=\arctan(a_2/a_1)$  (see Fig.~\ref{fig:lattices}{\bf (b)}). In analogy to $d=1$, we could restrict the configurations of spin coherent states  to be translationally-invariant within these sub-lattices $\{\theta(\boldsymbol{x}),\phi(\boldsymbol{x})\}\mapsto\{\theta_{s},\phi_{s}\}$, where  $s\in\{A,B\}$, and still account for  anti-ferromagnetic configurations when the respective $A,B$ angles differ, or for ferromagnetic ones when they are equal. However, this choice may not suffice to capture the groundstate ordering of the effective strong-coupling model~\eqref{eq:spin_model}. For illustrative purposes,  consider the limit $a_1\gg a_2$ and $r\approx 1$ so that, in light of the spin-spin couplings in Eq.~\eqref{eq:couplings_2d_wp},  $|J_{2,\mathsf{x}}|\gg \{|J_{j,\mathsf{a}}|, \forall j\neq 2, \mathsf{a}\neq\mathsf{x}\}$. Since this leading spin coupling is negative $J_{2,\mathsf{x}}<0$, the spins will want to align ferromagnetically along each of the columns, adopting polar and azimuthal angles $\theta_A^*=\theta_B^*=\pi/2$, $\phi_A^*=\phi_B^*\in\{0,\pi\}$. Additionally, since the perturbative coupling between neighbouring columns fulfils $J_{1,\mathsf{x}}>0$, spins in adjacent columns might minimise the groundstate energy by choosing opposing azimuthal angles $\phi_A^*=\phi_B^*\in\{\pi,0\}$. Since this is inconsistent with the $A,B$ sub-lattice layout, allowing for this possible ordering in the parametrisation of the constrained effective potential requires augmenting the number of configurations by considering a 4-site unit cell $\{\theta(\boldsymbol{x}),\phi(\boldsymbol{x})\}\mapsto\{\theta_{s,c},\phi_{s,c}\}$, where $c\in\{\mathsf{l},\mathsf{r}\}$ labels the left and right corners, as depicted in Fig.~\ref{fig:lattices}{\bf (b)}. By directly incorporating the non-linear constraint as we did forEq.~\eqref{eq:eff_potential}, the effective potential  reads in this case
\begin{widetext}
\beq
\frac{2V_{\rm eff}}{N_1N_2}\!=\!\sum_{c}\!\Big(\!\sum_{j}\sin\theta_{A,c}\sin\theta_{B,\tilde{c}(j)}\left(\tilde{J}_{j,\mathsf{x}}\cos\phi_{A,c}\cos\phi_{B,\tilde{c}(j)}+\tilde{J}_{j,\mathsf{y}}\sin\phi_{A,c}\sin\phi_{B,\tilde{c}(j)}\right)+\tilde{J}_{j,\mathsf{z}}\cos\theta_{A,c}\cos\theta_{B,\tilde{c}(j)}+\frac{h_{\mathsf{z}}}{2}(\cos\theta_{A,c}+\cos\theta_{B,c})\!\Big)\!,
\eeq
\end{widetext}
where we have introduced the function  $\tilde{c}(2)=c$ when the  spin-spin couplings occur along the $\boldsymbol{e}_2$ spatial direction, and  $\tilde{c}(1)=\overline{c}$ along  $\boldsymbol{e}_1$. For the latter, we define $\overline{c}=\mathsf{r (l)}$ for $c=\mathsf{l (r)}$, which swaps the left and right corners.   

The saddle-point conditions~\eqref{eq:saddle_point_conditions} corresponding to this potential lead to a non-linear system of 8 equations  which, once again, must  be solved numerically for generic cases. The exception is the standard limit $r=1$, where the effective spin model reduces to the 90$^{\rm o}$ compass model~\cite{PhysRevB.71.195120} in a transverse field, and one finds
\beq
\begin{split}
\phi_{s,c}^\star&\in\left\{\!\frac{\pi}{2},\frac{3\pi}{2}\!\right\},\hspace{1ex}\theta_{s,c}^\star=\pi-\arccos\left(\frac{h_{\mathsf{z}}}{|J_{1,\mathsf{y}}|}\right)\!\!, \hspace{1ex} {\rm if}\hspace{0.5ex} a_1<a_2,\\
\phi_{s,c}^\star&\in\left\{0,\pi\right\},\hspace{4.ex}\theta_{s,c}^\star=\pi-\arccos\left(\frac{h_{\mathsf{z}}}{|J_{2,\mathsf{x}}|}\right)\!\!, \hspace{1ex} {\rm if}\hspace{0.5ex}a_1>a_2.
\end{split}
\eeq
The SSB order parameters for these solutions are
\beq
\label{eq:solution_r_12}
\begin{split}
\langle\boldsymbol{S}(\boldsymbol{x})\rangle=\pm S\sqrt{1-\left(\frac{h_{\mathsf{z}}}{J_{1,\mathsf{y}}}\right)^{\!\!\!2}}\boldsymbol{e}_{\mathsf{y}}-S\frac{h_{\mathsf{z}}}{|J_{1,\mathsf{y}}|}\boldsymbol{e}_{\mathsf{z}},\hspace{1ex} {\rm if}\hspace{0.5ex} a_1<a_2,\\
\langle\boldsymbol{S}(\boldsymbol{x})\rangle=\pm S\sqrt{1-\left(\frac{h_{\mathsf{z}}}{J_{2,\mathsf{x}}}\right)^{\!\!\!2}}\boldsymbol{e}_{\mathsf{x}}-S\frac{h_{\mathsf{z}}}{|J_{2,\mathsf{x}}|}\boldsymbol{e}_{\mathsf{z}},\hspace{1ex} {\rm if}\hspace{0.5ex} a_1>a_2,
\end{split}
\eeq
which predict critical points at $|h_{\mathsf{z}}/J_{1,\mathsf{y}}|_{\rm c}=1$ if $a_1<a_2$, and $|h_{\mathsf{z}}/J_{2,\mathsf{x}}|_{\rm c}=1$ if $a_1>a_2$. Accordingly, for $|h_{\mathsf{z}}|<|J_{2,\mathsf{x}}|$ and a larger horizontal lattice spacing $a_1>a_2$, the SSB order parameter  $\langle S_{\mathsf{x}}(\boldsymbol{x})\rangle\propto \Pi_1$ corresponds to  ferromagnetic ordering along the internal $\mathsf{x}$ axis FM$_{\mathsf{x}}$, which corresponds to the inversion-breaking $\pi$ condensate of Eq.~\eqref{eq:pi_condensates} for the underlying four-Fermi model. Alternatively, for $|h_{\mathsf{z}}|<|J_{1,\mathsf{y}}|$ and a larger vertical lattice spacing $a_1<a_2$, the SSB order parameter describes ferromagnetic ordering along the internal $\mathsf{y}$ axis FM$_{\mathsf{y}}$, which corresponds to the other inversion-breaking $\pi$ condensate  $\langle S_{\mathsf{y}}(\boldsymbol{x})\rangle\propto \Pi_2$. We note  that these large-$S$ solutions for $r=1$ coincide with the variational mean-field estimates discussed in~\cite{ziegler2020correlated,ziegler2021largen}, and recall again that these  condensates are different from the Aoki parity-breaking phase.

\begin{figure}[t]
	\centering
	\includegraphics[width=0.5\textwidth]{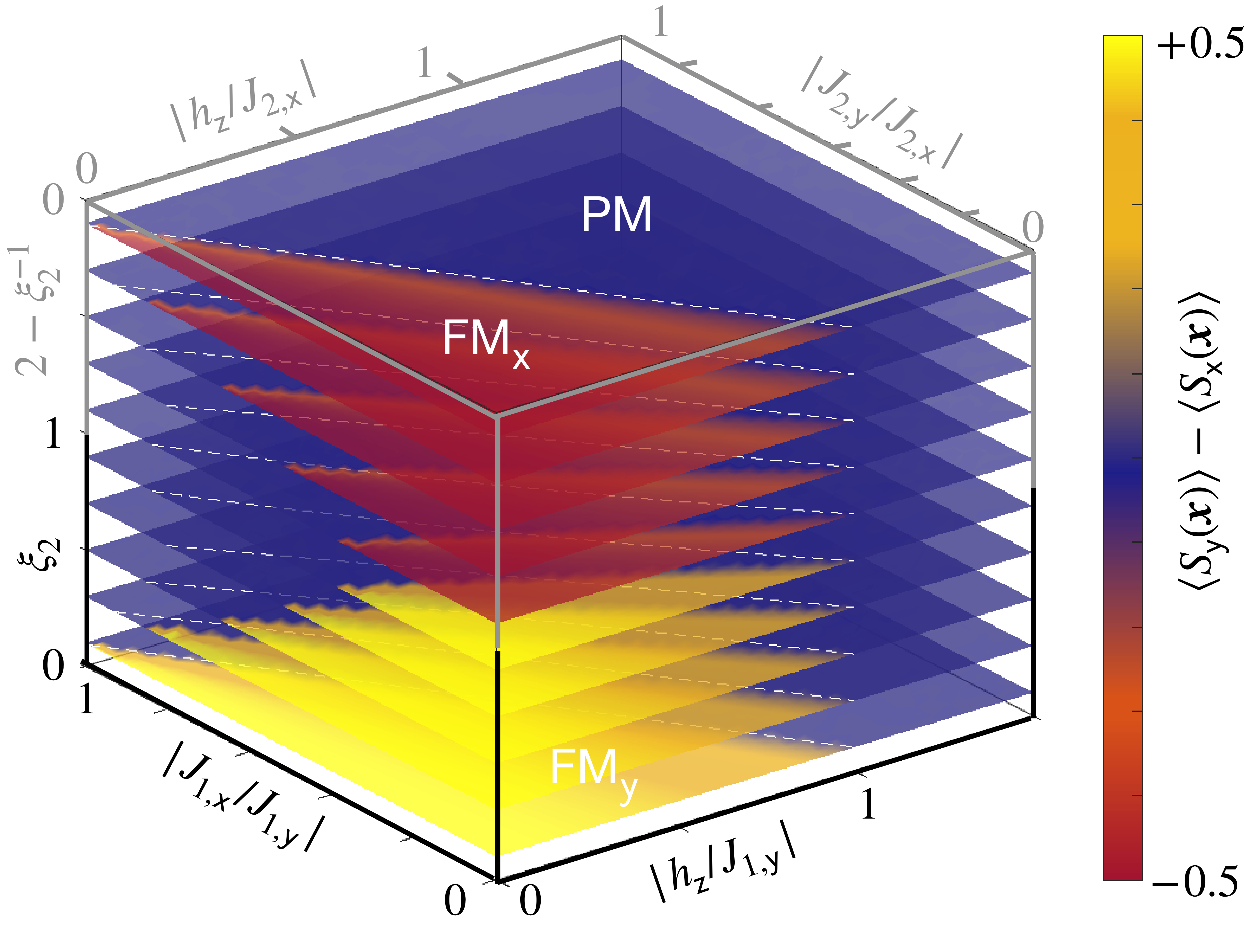}
	\caption{{\bf Large-$S$ compass magnetism   for $d=2$:} The large-$S$ method predicts that the type of ferromagnetic SSB order (inversion-breaking) condensate  corresponds to FM$_{\mathsf{x}}$ or FM$_{\mathsf{y}}$ for $\xi_2>1$ or $\xi_2<1$, respectively. To visualise the phase diagram in a single figure, we present stacked contour plots of the difference of the ferromagnetic order parameters $\langle S_{\mathsf{y}}(\boldsymbol{x})\rangle-\langle S_{\mathsf{x}}(\boldsymbol{x})\rangle$ as a function of the relative transverse field and the Wilson parameter via $|J_{1,\mathsf{x}}/J_{1,\mathsf{y}}|=|J_{2,\mathsf{y}}/J_{2,\mathsf{x}}|=(1-r^2)/(1+r^2)$.  The  contour plot shows FM$_{\mathsf{x}}$ in red scale, FM$_{\mathsf{y}}$ in yellow scale, and PM in blue scale. We also include    dashed lines for the large-$S$ analytical estimates~\eqref{eq:critical_lines_2d}.   }
	\label{fig:d_2_ssb}
\end{figure}

To treat $0<r<1$, we must solve the problem numerically. We use the same strategy as described for $d=1$, which combines a coarse global minimisation with more efficient non-linear programming methods that are consistently initialised to yield accurate estimates of the potential minima. We  obtain the SSB order parameters   from the numerical saddle points $\{\theta_{s,c}^\star,\phi_{s,c}^\star\}$ for various Wilson parameters $r\in\{0.5,0.6,0.7,0.8,0.9,1\}$. The corresponding magnetisations display similar non-analytic behaviours, which can be used to infer the location of the critical points, and how these flow as one varies $r$. In  Fig.~\ref{fig:d_2_ssb}, we present a stack of two-dimensional contour plots that summarises the large-$S$ phase diagram, and shows a clear dependence on  both the Wilson parameter and the  anisotropy parameter
\beq
\xi_2=\frac{a_1}{a_2}.
\eeq{}In this contour plot, we represent the difference of the two possible SSB order parameters $\langle{S}_\mathsf{y}(\boldsymbol{x})\rangle-\langle{S}_\mathsf{x}(\boldsymbol{x})\rangle$, such that negative (positive) values signal a FM$_\mathsf{x}$ (FM$_\mathsf{y}$) phase with a $\Pi_1$($\Pi_2$) Lorentz-breaking condensate, and are depicted in red (yellow) scale. In the stacking  $z$ direction, we plot the anisotropy parameter for $\xi_2\in[0,1]$ for $a_1<a_2$ (black axis), while we represent $2-1/\xi_2\in[1,0]$ for $a_2>a_1$ (grey axis). 

The lower stacked contour plots thus represent FM$_{\mathsf{y}}$ ordering, whereas the upper ones represent FM$_{\mathsf{x}}$. In the $x$ and $y$ axes of these contour plots, we select the relevant normalised couplings such that the stacked contour plots are completely symmetric as one crosses the isotropic configuration $a_1=a_2$. In general, one observes that the  critical point $|h_{\mathsf{z}}/J_{2,\mathsf{x}}|_{\rm c}=1$ ($|h_{\mathsf{z}}/J_{1,\mathsf{y}}|_{\rm c}=1$) at  $r=1$ and $\xi_2>1$ ($\xi_2<1$), changes as one decreases $r$,  increasing in this way the remaining coupling strengths of the compass Heisenberg model~\eqref{eq:couplings_2d_wp}, namely  $|J_{2,\mathsf{z}}/J_{2,\mathsf{x}}|=|J_{2,\mathsf{y}}/J_{2,\mathsf{x}}|$ ( $|J_{1,\mathsf{z}}/J_{1,\mathsf{y}}|=|J_{1,\mathsf{x}}/J_{1,\mathsf{y}}|$). In addition, one can also observe how the critical points change with anisotropy when $r<1$, such that the extent of the Lorentz-breaking fermion condensates in general depends on the anisotropy of the lattice regularisation.  To gain some further understanding, we can extend the previous discussion of the $d=1$ case around Eq.~\eqref{eq:mf_decoupling} to cover $d=2$ in the regime of very large anisotropy, since in this case the compass Heisenberg-Ising models reduce to very weakly-coupled columns (rows). To derive a self-consistent mean-field decoupling for $d=2$, note that we now have to deal with these additional spin-spin couplings between adjacent rows (columns) for $a_1> a_2$ ($a_1< a_2$). Setting $r\approx 1$, one can solve the corresponding self-consistent equations approximately, leading to 
\beq
\label{eq:critical_lines_2d}
\begin{split}
\left|\frac{h_{\mathsf{z}}}{J_{1,\mathsf{y}}}\right|_{\rm c}&\!\!\!\!\approx1-\frac{1-r^2+(1-r^2)\xi_2^2}{1+r^2}\frac{1-(1-r^2)\xi_2^2}{1+r^2},\hspace{1ex} {\rm if}\hspace{0.5ex} a_1<a_2,\\
\left|\frac{h_{\mathsf{z}}}{J_{2,\mathsf{x}}}\right|_{\rm c}&\!\!\!\!\approx1-\frac{1-r^2+(1-r^2)\xi_2^{-2}}{(1+r^2)}\frac{1-(1-r^2)\xi_2^{-2}}{1+r^2},\hspace{0.5ex} {\rm if}\hspace{0.5ex} a_1>a_2.
\end{split}
\eeq
The white dashed lines of Fig.~\ref{fig:d_2_ssb} represent these analytical predictions of the critical points. As can be observed, they match nicely the numerical critical lines for large anisotropies $\xi_2\ll1 $ ($\xi_2\gg 1$), but the discrepancy increases as one approaches the isotropic case $\xi_2=1$.

 \begin{figure}[t]
	\centering
	\includegraphics[width=0.4\textwidth]{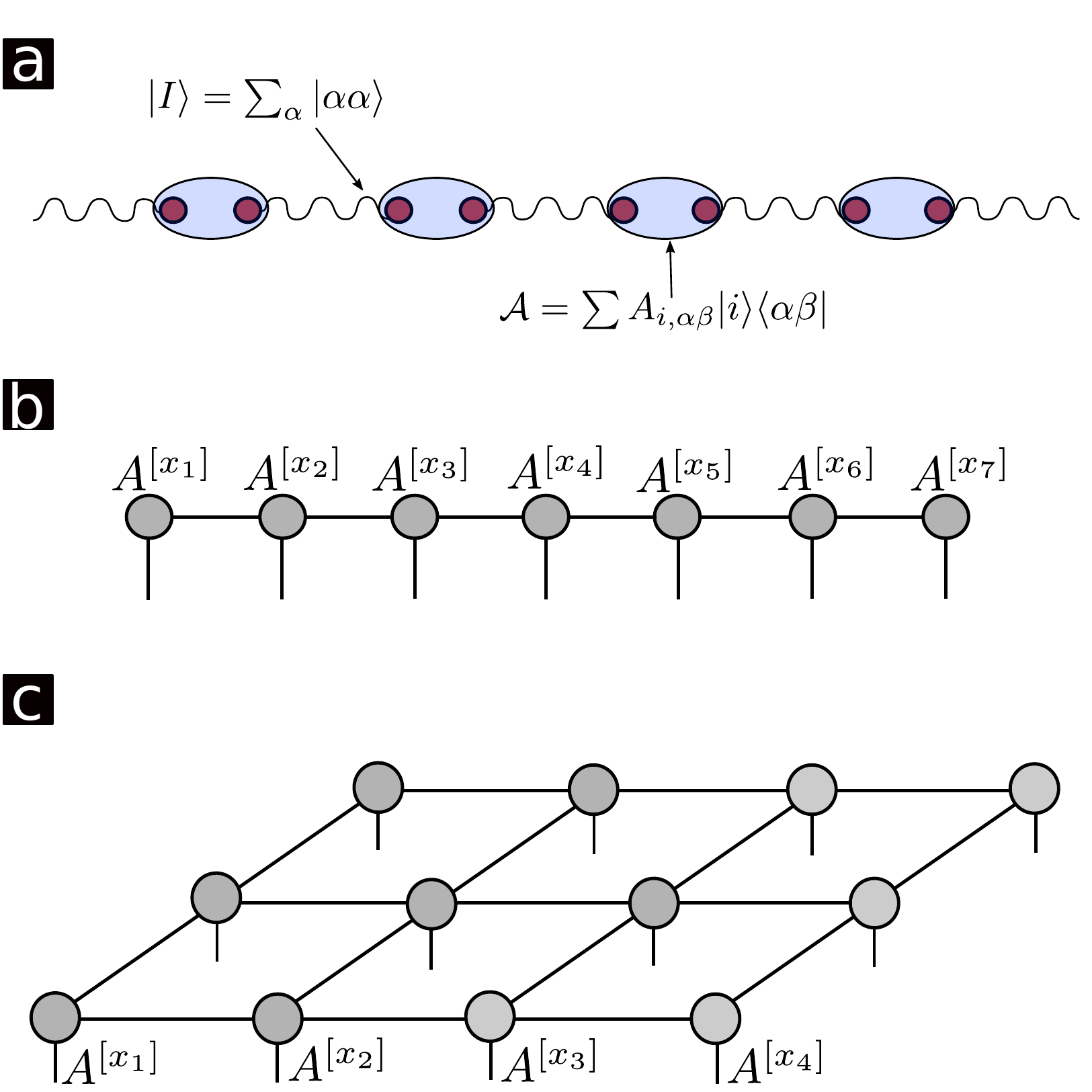}
	\caption{{\bf Tensor network representations:} {\bf (a)} Matrix product states as a network of maximally entangled states $|I\rangle$ shared between physical sites of the one dimensional lattice to which local operations $A_j$ are applied on combined virtual space on each site. {\bf (b)} Diagrammatic representation of MPS characterised by a three leg tensor $A$ defined in every site throughout the tensor network.
	{\bf (c)} Diagrammatic representation of PEPS corresponding to square lattice. }
	\label{fig:tn_sketch}
\end{figure}

\section{\bf Tensor-Network numerical simulations}
\label{sec:tn}
In this section, we test the validity of the  large-$S$ predictions   of  Sec. \ref{sec:eff_spin_model} by means of variational algorithms based on tensor network states (TNSs) \cite{verstraete2008matrix,orus2014practical,ran2020tensor}. 

The quantum state of a lattice model composed of $N_{\rm s}$ $\tilde{d}$-level systems, i.e. spins,  can be written in the basis
of  tensor products of local states. The quantum state is then fully characterised by the coefficients of 
these basis states, which are tensors $C_{i_1,i_2, \cdots, i_N}$ of rank $N_{\rm s}$ and dimension $\tilde{d}$
\beq 
| \psi \rangle = \sum_{i_1,i_2, \cdots, i_{N_{\rm s}}} C_{i_1,i_2, \cdots, i_{N_{\rm s}}} |i_1,i_2, \cdots, i_{N_{\rm s}} \rangle .
\eeq
Here, we have introduced the indexes $i_{n}\in\{1,\cdots,\tilde{d}\}$,  such that the description of the state requires of $\tilde{d}^{N_{\rm s}}$ complex parameters. This
exponential growth makes this generic description
 unsuitable for numerical analysis. However, in a number of situations, the physically-relevant many-body states admit 
 a more concise description based on TNSs~\cite{cirac2009renormalization,evenbly2014algorithms}. Obtained  from a contraction of low-rank tensors on so-called virtual 
indices, TNSs economically approximate the states of a system with local
interactions in thermal equilibrium. The number of required parameters scales only polynomially with system size \cite{molnar2015approximating}, circumventing the previous exponential growth of the most generic description. In fact, these variational states are based on the powerful insights related to the area law \cite{eisert2008area,eisert2013entanglement}. The area law places bounds on quantum entanglement that a many-body system can generate, which translates directly to the number of parameters required to describe a physically-relevant quantum state.

Tensor-network calculations benefited from the advent of White's density matrix renormalisation group (DMRG) \cite{white1992density}, famous for its extraordinary accuracy in solving one-dimensional quantum systems,  which is intimately connected with a tensor decomposition known as the matrix product state (MPS) \cite{rommer1997class,Dukelsky_1998}. 
These variational states can be understood in terms of pairs of maximally-entangled states on neighbouring lattice sites, which describe auxiliary degrees of freedom, and get locally projected onto the lower-dimensional subspace of physical spins  at each lattice site.
In fact, a very useful and intuitive way of thinking about MPS is the following valence-bond construction. Consider $N_{\rm s}$ spins aligned on a ring, the states of which are labelled by the internal index $i$. One   assigns two auxiliary spins of dimension $D$ to each of these physical spins,  assuming that each pair of neighbouring  auxiliary spins is initially in a maximally entangled state $|I\rangle=\sum_{\alpha=1}^D |\alpha,\alpha\rangle$, often referred to as an entangled bond. Applying the map that plays the role of the aforementioned projector
\beq
\mathcal{A} = \sum A_{i,\alpha \beta} |i\rangle \langle \alpha \beta |
\eeq
to each of the $N_{\rm s}$ spins, and interpreting $A_i$ as a $D \otimes D$ matrix, we find that the coefficients of the final state can be expressed by a matrix product
${\rm Tr} \left[A_{i_1} A_{i_2} \cdots A_{i_N} \right]$ (see Fig. \ref{fig:tn_sketch} {\bf (a)}). In the $d=1$-dimensional models discussed in this work, we consider physical spins $S=1/2$, such that $i_n=s_{\boldsymbol{x}_n}\in\{\uparrow,\downarrow\}$, $\tilde{d}=2$, and $n \in \left \lbrace 1, \cdots, N_{\rm s} \right \rbrace$ with $N_{\rm s}=N_1$ being the number of sites of the spatial chain. In 
general, the 
dimension of the entangled state $|I \rangle$ can be site-dependent and 
we write $A^{[\boldsymbol{x}_n]}_{s_{\boldsymbol{x}_n}}$ for the $D_n \times D_{n+1}$ matrix corresponding to site
$\boldsymbol{x}_n$; the states
then have the form
\beq 
|\psi \rangle = \sum_{\{s_{\boldsymbol{x}_n}\}} {\rm Tr} \left \lbrace A^{[{s_{\boldsymbol{x}_1}}]}_{s_{\boldsymbol{x}_1}} A^{[{s_{\boldsymbol{x}_2}}]}_{{s_{\boldsymbol{x}_2}}} \cdots A^{[{s_{\boldsymbol{x}_{N_{\rm s}}}}]}_{{s_{\boldsymbol{x}_{N_{\rm s}}}}} \right \rbrace |{s_{\boldsymbol{x}_1}},{s_{\boldsymbol{x}_2}},\cdots, {s_{\boldsymbol{x}_{N_{\rm s}}}}\rangle
\eeq
and are called MPS.
This construction can be mathematically expressed as a network of tensors with multiple indexes corresponding to the physical and auxiliary degrees of freedom, such that those corresponding to the auxiliary ones are contracted as described in Fig. \ref{fig:tn_sketch} {\bf (b)}. In this case the number of parameters needed to describe a physical state in the MPS language scales as $O(N_1 d D^2)$ with $d$ the physical dimension of the spins.

A natural generalization of MPS to two, or even higher, spatial dimensions is represented by projected entangled pair states (PEPS \cite{murg2007variational}). Again, this kind of state can be understood 
in terms of pairs of maximally-entangled states of neighbouring auxiliary systems, which are
are locally projected into the low-dimensional physical subspace.
As represented in Fig. \ref{fig:tn_sketch} {\bf (c)}, the PEPS describes a state through interconnected tensors.  
For the  two-dimensional spatial lattices considered in this work, which consist of $N_{\rm s}=N_1N_2$ sites, we specify  the PEPS variational ansatz~\cite{verstraete2004renormalization,murg2007variational,cirac2021matrix} as 
\beq \label{eq:def_peps}
|\psi \rangle = \!\!\!\sum_{ s_{\boldsymbol{x}_n}}\!\!\! F\!\!\left(\! A^{[\boldsymbol{x}_1]}_{s_{\boldsymbol{x}_1}},A^{[\boldsymbol{x}_2]}_{s_{\boldsymbol{x}_2}}, \cdots, A^{[\boldsymbol{x}_{N_1\!N_2}]}_{s_{\boldsymbol{x}_{N_1\!N_2}}} \right)
|s_{\boldsymbol{x}_1},s_{\boldsymbol{x}_2}, \cdots, s_{\boldsymbol{x}_{N_1\!N_2}} \rangle.
\eeq
This PEPS is represented by a network of  $N_1N_2$ tensors $A^{[\boldsymbol{x}_m]}_{s_{\boldsymbol{x}_n}}$, some of which are connected  according to the geometry of the lattice and the notion of neighbouring lattice sites.
Each tensor of the PEPS has $N_{\rm b}$ so-called bond indices of dimension $D_n$, which describe the aforementioned auxiliary degrees of freedom, and a single physical index of dimension $d$. The choice of $N_{\rm b}$ in the tensor network can be arbitrary, and typically depends on the geometry of the lattice.
For example for a $N_1 \times N_2$ lattice, a PEPS contains $N_1 \times N_2$ bulk tensors with $N_{\rm b}=4$ and $D_n=D$. Each tensor depends on $d D^4$
complex coefficients. Therefore the PEPS is characterized by $O(N_s d D^4)$ parameters.
The function $F$ contracts all the tensors $A^{[\boldsymbol{x}_m]}_{s_{\boldsymbol{x}_m}}$, according to this pattern, and then performs the trace to obtain a scalar quantity such that Eq.~\eqref{eq:def_peps} can be understood as a parametrisation of a  particular set of states in the exponentially-large physical Hilbert space.

In this manuscript, we study the groundstate properties of quantum lattice Hamiltonians using different strategies for $d=1$  and $d=2$ spatial dimensions.
For $d=1$ we variationally optimize the MPS tensors, 
so as to minimise the expectation value of the corresponding Hamiltonian. 
By contrast for $d=2$, in analogy to spectroscopic methods that determine the particle spectrum via the imaginary-time evolution of correlators
in Euclidean LFTs \cite{gattringer_lang_2010}, we evolve the system in imaginary time until a stationary state corresponding to the groundstate is reached. This assumes that this groundstate is unique, and that the energy gap is non-zero, as done in the time-evolving block-decimation method (TEBD) for one-dimensional chains \cite{vidal2007classical,orus2008infinite}. In the following, we will use this method in the thermodynamic limit for the infinite PEPS state (iPEPS) \cite{orus2009simulation,corboz2011stripes}.

\begin{figure}[t]
	\centering
	\includegraphics[width=0.45\textwidth]{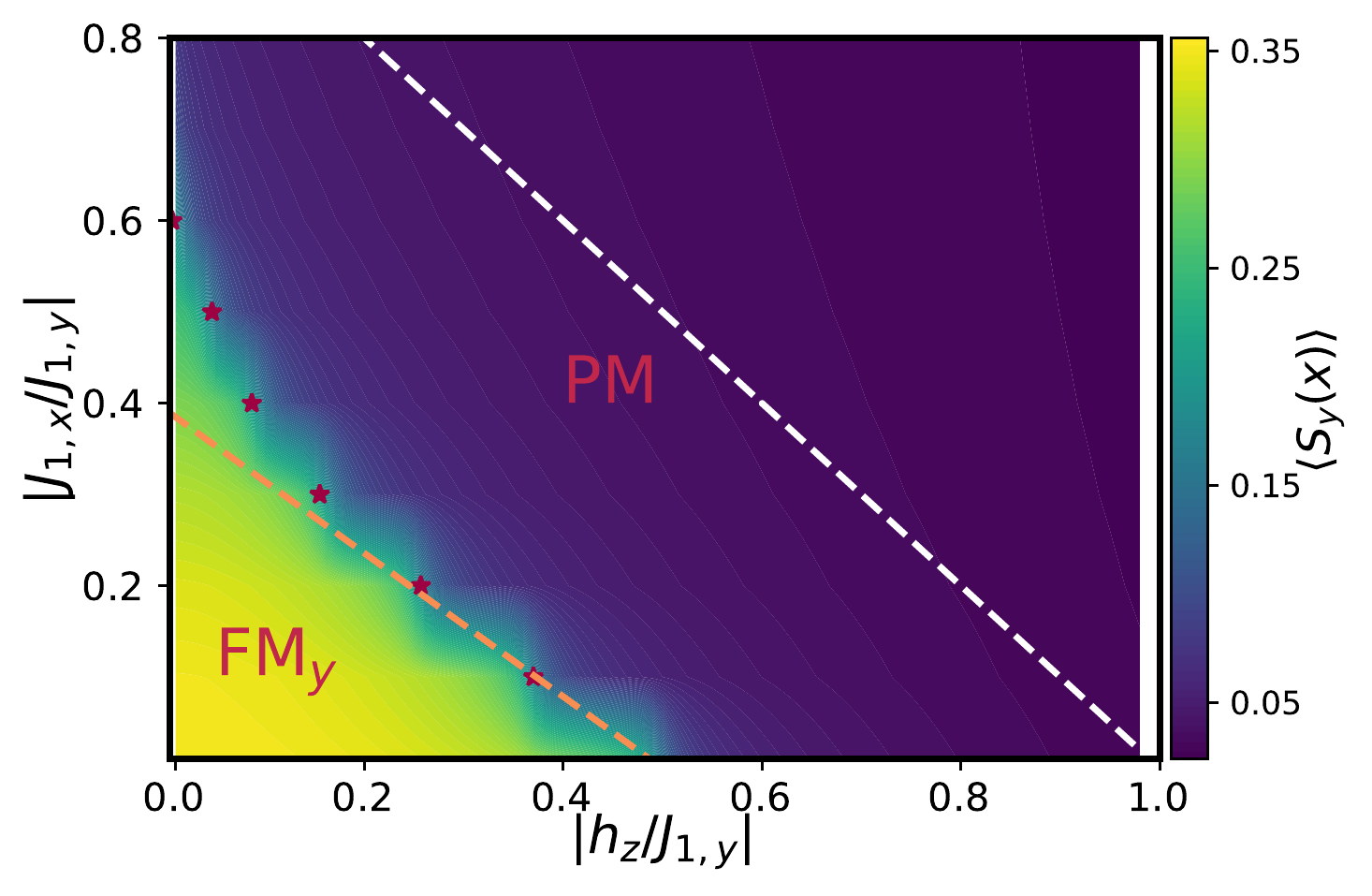} 
	\caption{{\bf Phase diagram of Heisenberg-Ising chain:} The  phase  diagram  display  two  regions  hosting a long-range-ordered  ferromagnetic  phase  (FM$_y$), and a paramagnetic phase (PM). The horizontal axis represents the magnetic field $h_z$, where as the vertical axis corresponds  to  the  ratio  of  the  tunnelling strengths $J_{1,z}/J_{1,y}$. The red stars (yellow dashed lines) show the critical points found from DMRG (self-consistent mean-field) numerics.  These points are plotted on top of the contour plot of the magnetisation $\langle S_y(x) \rangle$ obtained using DMRG.   }
	\label{fig:phase_diagram_mps_1d}
\end{figure}

\begin{figure}[t]
	\centering
	\includegraphics[width=0.4\textwidth]{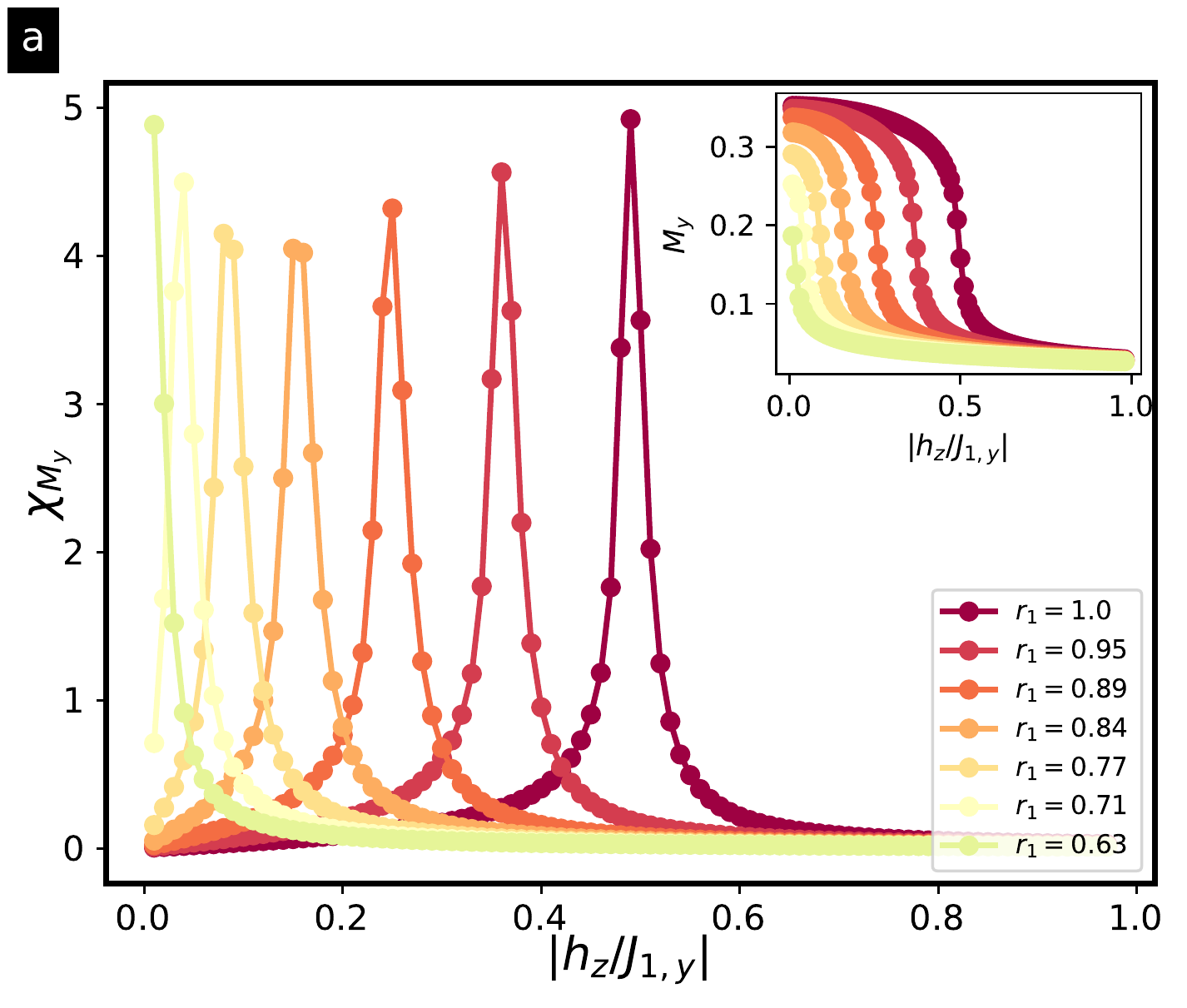} \\
	\includegraphics[width=0.4\textwidth]{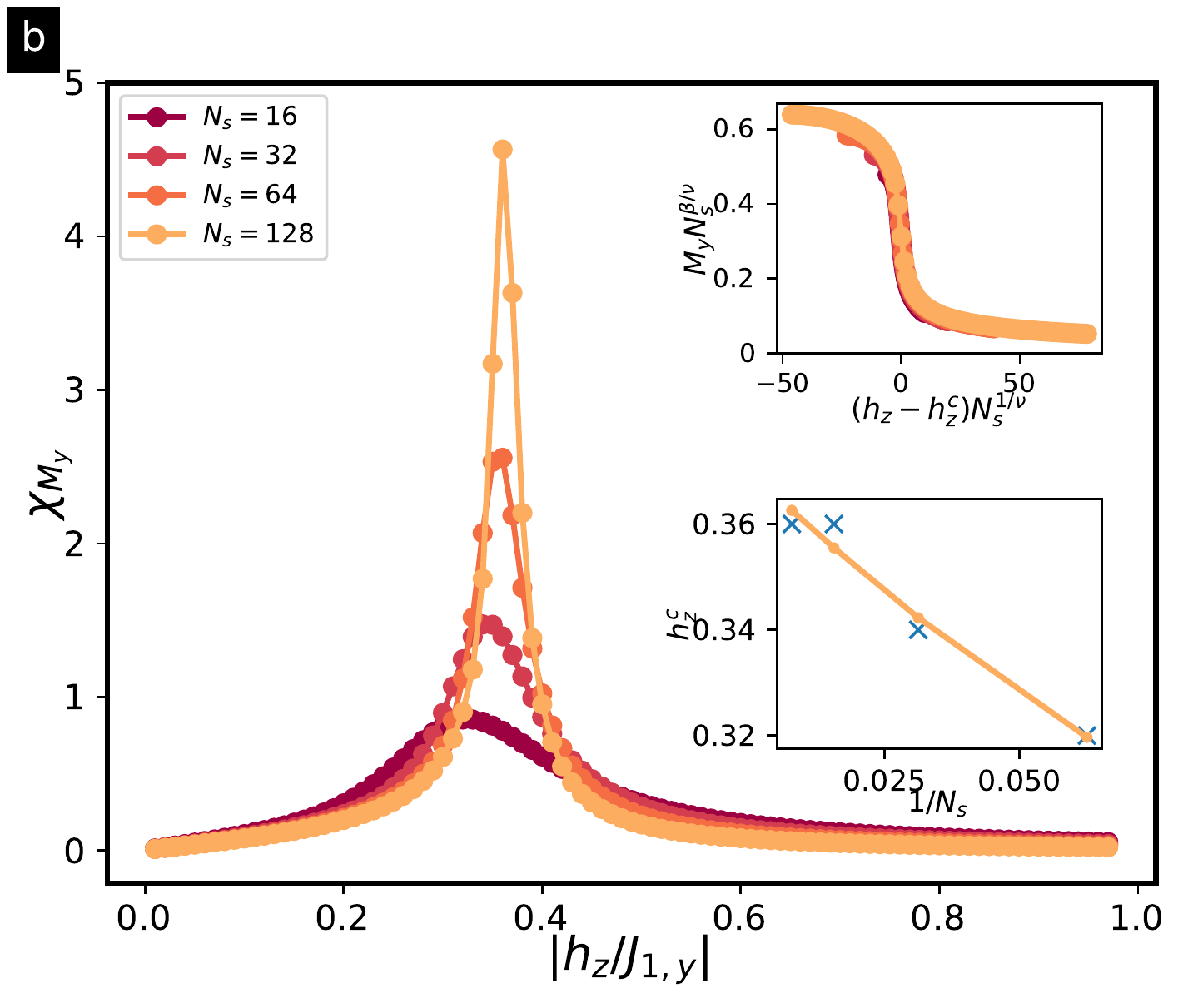}
	\caption{{\bf Ferromagnetic and paramagnetic susceptibilities:} {\bf (a)} The ferromagnetic susceptibility $\chi_{M_y}$, for fixed coupling strength $J_y =-1$, and for different couplings $J_x$. As magnetic field $h_z$ is varied it develops peaks at the critical points. In the inset, we show ferromagnetic magnetization along the $y$ direction. The system develops a non-zero expectation value for transverse fields below a critical value $h_z < h^c_z$. {\bf (b)} The paramagnetic susceptibility $\chi_{M_z}$, for the same parameters, which develops peaks at those critical points. In the inset, we show ferromagnetic magnetisation along the $z$ direction.}
	\label{fig:susceptibility_mag_1d}
\end{figure}


\subsection{Tensor-network Ising magnetism for $d=1$}
\label{sec:compass_mps_d_1}
In this section, we analyse the effect of correlations in the
phase diagram of the Heisenberg-Ising chain~\eqref{eq:spin_model} with spin-spin couplings  defined in Eq.~(\ref{eq:couplings_1d_wp}), and subjected to an additional transverse field in Eq.~\eqref{eq:transverse_field}. All of these parameters depend on  the Wilson $r$, and the goal is to explore the phase diagram as it is varied within $0<r<1$. In particular, we benchmark the large-$S$ results 
discussed in  Sec.~\ref{sec:largeS_d1},  giving more accurate predictions of the phase diagram and critical points presented in  Fig. \ref{fig:d_1_ssb} {\bf (c)}. 

As discussed in Sec.~\ref{sec:largeS_d1}, the Heisenberg-Ising chain  presents a critical line separating the ferromagnet FM$_{\mathsf{y}}$ and the paramagnet PM. 
In Fig. \ref{fig:phase_diagram_mps_1d}  we present the corresponding  MPS phase diagram as a function of relative coupling strengths $|J_{1,\mathsf{x}}/J_{1,\mathsf{y}}|=|J_{1,\mathsf{z}}/J_{1,\mathsf{y}}|$ and $|h_{\mathsf{z}}/J_{1,\mathsf{y}}|$. 
Our numerical results for the phases of matter are extrapolated using the quasi-exact DMRG algorithm, as discussed in detail below. The lines represent the critical points where the
SSB phase transitions occur, either obtained with  DMRG based on finite MPS with bond dimension $D=200$ (red stars), or by self-consistent mean-field method
 (yellow dashed lines), which exploits exact solutions of the transverse-field Ising model to derive a self-consistent equation for the transverse magnetization that can be solved analytically in the limit $|J_{1,\mathsf{x}}/J_{1,\mathsf{y}}|=|J_{1,\mathsf{z}}/J_{1,\mathsf{y}}|\ll 1$. This yields
 \beq
\label{eq:Ising_prediction}
\left|\frac{h_{\mathsf{z}}}{J_{1,\mathsf{y}}}\right|_{\rm c}=\frac{1}{2}-\frac{4}{\pi}\frac{1-r^2}{1+r^2},
 \eeq
 which must be compared to the large-$S$ estimate~\eqref{eq:Ising_prediction}, depicted with a white dashed line in the figure.
 
As can be observed in  Fig. \ref{fig:phase_diagram_mps_1d}, for small relative coupling $|J_{1,{\rm x}}/J_{1,{\rm y}}|$, attained for $r\approx 1$, the self-consistent mean-field and DMRG critical points separating the FM$_y$ and PM regions yield two critical lines that are very similar. Increasing $|J_{1,\mathsf{x}}/J_{1,\mathsf{y}}|$, larger differences appear between the critical lines, since  the self-consistent mean-field predicts a smaller FM$_y$ region. Note, however, that this prediction is strictly valid only in the vicinity of the exact critical point $\left|{h_{\mathsf{z}}}/{J_{1,\mathsf{y}}}\right|_{\rm c}={1}/{2}$. In comparison with the large-$S$ white-dashed line, we see that there is a large quantitative difference of the critical lines, e.g. for $r=1$ $\left|{h_{\mathsf{z}}}/{J_{1,\mathsf{y}}}\right|_{\rm c}=1$ within large-$S$, whereas $\left|{h_{\mathsf{z}}}/{J_{1,\mathsf{y}}}\right|_{\rm c}={1}/{2}$ with DMRG. This difference is characteristic of large-$S$ methods, and is a consequence of how we neglect quantum fluctuations by taking the saddle-point solution. However, note that the large-$S$ captures the physics of the model qualitatively correctly: it predicts that the FM$_{\mathsf{y}}$ region (pseudo-scalar condensate) shrinks proportionally to the combination $(1-r^2)/(1+r^2)$, whenever $0<r<1$. Had we solved the lattice model for increasing $S$ using DMRG, we would have found that the two lines approach each other as the spin is increased $S\in\{1/2,3/2,5/2,\cdots\}$.  

In the background of Fig.  \ref{fig:phase_diagram_mps_1d}, we present a contour plot of the SSB order parameter, which corresponds to the pseudo-scalar condensate $\langle S_y(\boldsymbol{x})  \rangle \propto \Pi_5$ defined in Eq.(\ref{eq:pseudo-sclalr_condensate}). In order to avoid numerical problems due to the incomplete symmetry breaking of the magnetization $M_{\mathsf{y}}=\langle S_y(\boldsymbol{x})  \rangle =\frac{1}{N_1} \sum_{\boldsymbol{x}\in\Lambda_1} \langle S_{\mathsf{y }}(\boldsymbol{x})\rangle$, we determine instead the corresponding structure factors
\beq 
\begin{split}
\label{eq:structure_factors}
S_{\mathsf{y y}}(k) = \frac{1}{N_{\rm 1}^2} \sum_{n_1,n_1'} \ee^{\ii ka_1 (n_1-n_1')} \langle S_{\mathsf{y}}(n_1) S_{\mathsf{y}}(n_1')\rangle.
\end{split}
\eeq
The zero-momentum component of these structure factors yield the desired magnetisation in the thermodynamic limit $M_{\mathsf{y}}=\left( S_{\mathsf{yy}} (0) \right)^{1/2}$.
The contour plot of the magnetisation clearly identifies the SSB region in yellow-green scale with a non zero pseudo-scalar condensate, namely a Ising ferromagnet FM$_y$, which is separated from a region in blue scale, where the parity is preserved, namely a paramagnet PM with zero magnetization along the internal $\mathsf{y}$ axis. 

Let us now give some more details on the methodology used to extract numerically the critical points shown in  Fig. \ref{fig:phase_diagram_mps_1d}. 
By calculating the ferromagnetic magnetisations and, particularly,   the corresponding susceptibilities, we can identify the critical points occurring at a  non-zero external field $h_{\mathsf{z}} > 0$. In Fig. \ref{fig:susceptibility_mag_1d} {\bf (a)}, we present the susceptibility
$\chi_{\mathsf{y}}=\partial M_{\mathsf{y}} / \partial h_z$ for different values of Wilson parameter $r$, which clearly peaks at specific values of the ratio $\left|{h_{\mathsf{z}}}/{J_{1,\mathsf{y}}}\right|$ that move to the left as $r$ is decreased. In Fig. \ref{fig:susceptibility_mag_1d} {\bf (b)}, we display the ferromagnetic susceptibility $\chi_{\mathsf{y}}$ for different number of sites $N_s$, fixing $r=0.82$, and varying the external magnetic field $h_{\mathsf{z}}$.
The finite size scaling (FSS) of the magnetic susceptibility maxima, as a function of $N_1$ is displayed in the lower inset. As one can see in the inset, the peak of the chiral susceptibility at transverse field $h^c_{\mathsf{z}}$ diverges with the size of the chain, and fitting the maxima of $h^c_{\mathsf{z}}$ to $h^c_{\mathsf{z}}(N_1) = h^c_{\mathsf{z}}(1+a N^{-1}_1 +b N^{-2}_1 )$, we can delimit the ferromagnetic region and locate the phase transitions in the thermodynamic limit $N_s\to\infty$. Once the critical point is known, in the upper inset we show the data collapse of the  magnetization curves when rescaled with the system size using the critical exponent  of the 2D Ising universality class. Accordingly, the whole critical line delimiting  the Aoki phase belongs to this universality class, in spite of having perturbation to the Ising limit in the form of the additional spin-spin couplings~\eqref{eq:couplings_1d_wp} when $r<1$.

\begin{figure}[t]
	\centering
	\includegraphics[width=0.45\textwidth]{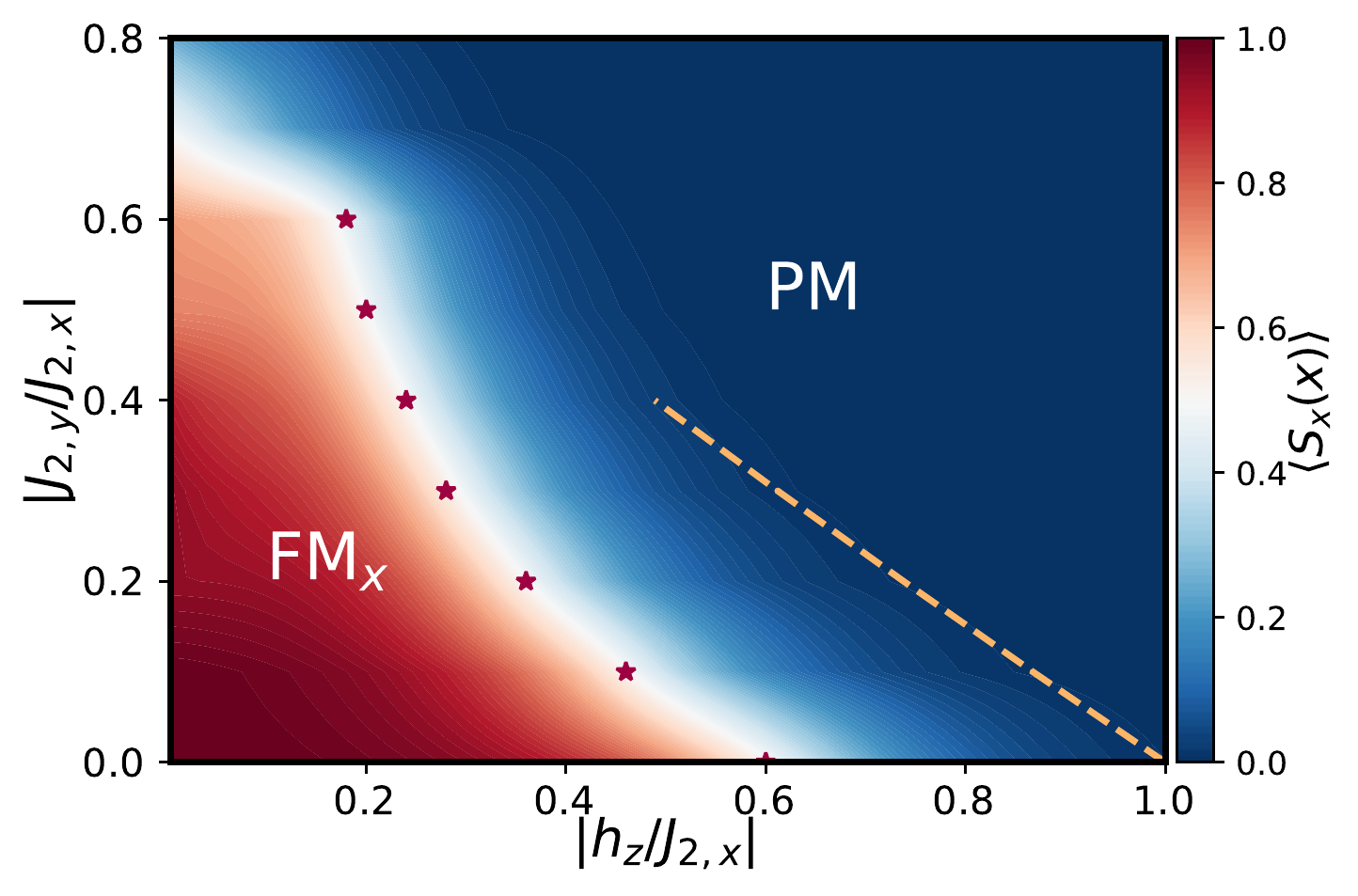} 
	\caption{{\bf Phase diagram of Compass Heisenberg-Ising model:} The  phase  diagram  display  two  regions  hosting a long-range-ordered  ferromagnetic phase  (FM$_x$), and a paramagnetic phase (PM). The horizontal axis represents the magnetic field $h_z$, where as the vertical axis corresponds to  the  ratio  of  the  tunnelling strengths $J_{2,y}/J_{2,x}$. The red stars (yellow dashed lines) show the critical points found from iPEPS algorithm (large-S predictions) numerics.  These points are plotted on top of the contour plot of the magnetization $\langle S_x(x) \rangle$, using the iTEBD algorithm for iPEPS ansatz. }
	\label{fig:phasediag_2d}
\end{figure}

\subsection{Tensor-network compass magnetism for $d=2$}
\label{sec:compass_peps_d2}

In this section, we show the results obtained by using the above iPEPS algorithm for the Heisenberg-Ising compass
model~\eqref{eq:spin_model} with spin-spin couplings defined in Eq.~(\ref{eq:couplings_2d_wp}), and subject to an additional transverse field in Eq.~\eqref{eq:t_f_2d}, working directly in the 
infinite-lattice limit. In particular, we have computed the ground state wave 
function $|\psi_{\rm GS}\rangle$ of the system by performing the imaginary-time 
evolution for different values of the spin couplings $\{J_{1,\mathsf {a}}$, $J_{2,\mathsf{a}}\}_{\mathsf{a}\in\{\mathsf{x,y,z}\}}$, and the 
transverse magnetic field $h_{\mathsf{z}}$, and then evaluated observable quantities on it, 
such as the groundstate energy and the local order parameters related to the 
 ferromagnetic phases.

In Sec.~\ref{sec:2d_largeS}, we used a large-$S$ method to predict a critical line separating the symmetry-broken  ferromagnets FM$_x$ and FM$_y$ from a paramagnet via second-order phase transitions. Under certain approximations, these critical lines can be analytically found~\eqref{eq:critical_lines_2d}, corresponding to the white dashed  lines of Fig. \ref{fig:d_2_ssb}. In order to test the validity of these large-$S$ predictions, we use our iPEPS algorithm for $D = 2$.
By measuring paramagnetic and ferromagnetic magnetisations, we confirm that these quantities can be used to identify the critical points also for a non-zero magnetic field $h_{\mathsf{z}}>0$.

\begin{figure}[t]
	\centering
	\includegraphics[width=0.4\textwidth]{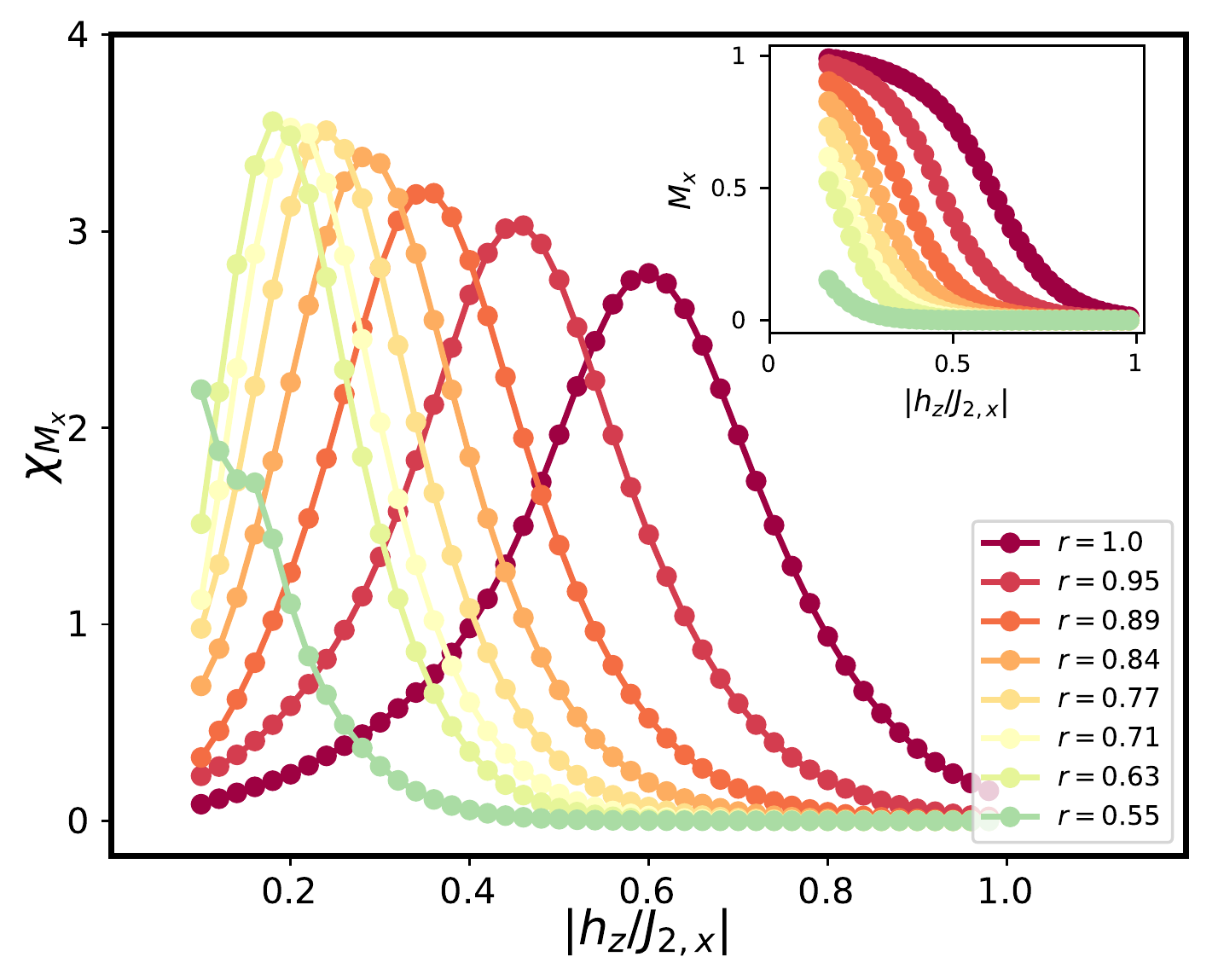}
	\caption{{\bf Ferromagnetic susceptibility:} The ferromagnetic susceptibility $\chi_{M_x}$  for $J_{2,y}=-1.0$ and $J_{1,x}=0.2$ shows a peak for different Wilson parameter $r$, and allow us to locate the critical points. In the upper inset we show the magnetization $M_x$ versus $h_z$. }
	\label{fig:susceptibilities_2d}
\end{figure}

We start by setting the spatial anisotropy to $\xi_2=3.16$, which would correspond to a specific stacked plane  in the large-$S$ phase diagram of Fig.~\ref{fig:d_2_ssb}, where FM$_{\mathsf{x}}$ order competes with the disordered PM. Our numerical iPEPs results for this competition are presented in Fig.~\ref{fig:phasediag_2d}. 
The lines correspond to the critical points where the FM$_{\mathsf{x}}$ and PM phase transitions occur, either obtained with a imaginary time evolution based on infinite iPEPS with bond dimension $D=2$ (red stars), or by the large-S approximate prediction~\eqref{eq:critical_lines_2d} (orange dashed lines). We observe a similar trend as in the case of $d=1$; the region of the inversion-breaking condensate shrinks as the value of the Wilson parameter $r$ is reduced within $0<r<1$. In the vicinity of the standard choice $r=1$, and for $\xi_2\gg1$, we see once again that the region of non-vanishing condensate decreases proportionally to the ratio $(1-r^2)/(1+r^2))=|J_{2,y}/J_{2,x}|$. On the other hand, as $r\to 0$, the iPEPs critical line bends upwards, as also occurred for $d=1$ (see the red stars in Fig.~\ref{fig:phase_diagram_mps_1d}). Note that, although the large-$S$ and iPEPs critical lines cross for smaller  values of $r$, i.e. larger ratios of $|J_{2,\mathsf{y}}/J_{2,\mathsf{x}}|$, the analytical predictions~\eqref{eq:critical_lines_2d}  are not strictly valid in this regime. On the other hand, for the regime where $r\approx 1$, we see that the large-$S$ predictions are closer to the iPEPs results in comparison to the $d=1$ case, showcasing that mean-field predictions typically improve as dimensionality increases.

Let us again discuss details on how we extracted such critical points numerically. 
In the inset of Fig.~\ref{fig:susceptibilities_2d}, we present the magnetisation $M_{\mathsf{x}}=\langle S_{\mathsf{x}}(\boldsymbol{x}) \rangle$
as a function of transverse magnetic field $h_\mathsf{z}$,
setting $J_{1,\mathsf{x}} = -1$ and exploring different values of $J_{1,\mathsf{y}} < 1$. This
figure shows that, for weak transverse fields, the magnetization
attains a non-zero value  signalling the broken symmetry
FM$_{\mathsf{x}}$ phase, which corresponds to the inversion symmetry-broken fermion condensate. The main panel shows the corresponding magnetic
susceptibility $\chi_{\mathsf{x}}$ peaking at a specific value of the
transverse field, which can be used to locate the corresponding
critical points. Note that these peaks are not as pronounced as for $d=1$ and that, given that we work with translationally-invariant iPEPS, we cannot perform FSS to see how the peak diverges and extract accurate estimations of the critical point and universality class. In future studies, it would be interesting to push the numerics to explore larger values of bond dimension $D> 2$, which would permit a more accurate location of the critical points. This question is particularly relevant in the regime $r\to 0$. where $d=1$ and $d=2$ results seem to differ qualitatively. In the 1D case, the Aoki phase shrinks all the way to zero, whereas in the 2D case it seems to survive. This could be related to the fact that the $r=0$ limit maps onto a Heisenberg model on a rectangular lattice, and that this model has long-range order in contrast to the 1D version. We note that these questions could also  be addressed with other  methods such as Monte Carlo simulation.

\section{\bf Conclusions and outlook}

In this work, we have explored the limit of strong Hubbard interactions in models of correlated topological insulators that arise for spin-orbit coupled fermions in lattices of one and two spatial dimensions. These models can be understood as the single-flavour limit $N=1$ of four-Fermi quantum field theories with a Wilson-type discretization, making an interesting and fruitful connection between condensed matter and lattice field theories. As discussed in this work, most lattice field theory studies fix the Wilson parameter of this discretization to $r=1$, as a non-unity value has trivial consequences in the non-interacting limit. However, understanding the role of $0<r<1$ in the presence of interactions is not clear {\it a priori}, which is the question explored in this work. Moreover, given the fact that these four-Fermi field theories are amenable to study using cold-atom quantum simulators, with spin-orbit coupled fermions in Raman lattices where the effective value of $r$ depends on  intensities of the lasers that control the Raman lattice, which are generically different, i.e. $r\neq 1$, the question addressed in this work is also relevant for understanding the possible phases that can be explored in possible cold-atom realisations. 

To address this question, we have derived an effective spin model in the limit of strong four-Fermi interactions, finding a specific dependence of the couplings on the Wilson parameter $r$. In $d=1$, the resulting model can be related to an $\mathsf{XXZ}$, also known as Heisenberg-Ising,  model in a staggered magnetic field, whereas in $d=2$ it is related to a compass model with directional spin-spin couplings, each of which is described by  different Heisenberg-Ising couplings. We have formulated a path-integral representation of the partition function, which connects the strongly-interacting limit with a constrained QFT: a non-linear sigma model with a discrete $\mathbb{Z}_2$ symmetry. This permits exact solutions in the large-$S$ limit, which enable us to identify the relevant phases of matter, and draw specific predictions about phase transitions and the flow of the critical points with the Wilson parameter $r$. The validity of these predictions have been tested against tensor-network numerical simulations, which show that the large-$S$ diagrams are qualitatively correct. On the other hand, the  numerical results give more accurate estimates of the flow of the critical lines and, in some cases, allow us to infer the correct universality class of the lines that differs from the large-$S$ mean-field-type scaling.

As an outlook, we believe that the present manuscript, together with  other recent works~\cite{PhysRevX.7.031057,BERMUDEZ2018149,tirrito2021topological,ziegler2020correlated,ziegler2021largen},  provide a rich 
cross-disciplinary toolbox  to understand interaction effects in topological matter. It will be interesting to exploit, adapt, and combine all of these different  techniques to study harder problems such as, for instance, abandoning  half-filling
and  exploring correlated topological phases at non-zero fermion densities.

\acknowledgements
A.B. acknowledges support
from  PGC2018-099169-B-I00 (MCIU/AEI/FEDER, UE), from
the Grant IFT Centro de
Excelencia Severo Ochoa  CEX2020-001007-S,
funded by MCIN/AEI/10.13039/501100011033, from the grant  QUITEMAD+
S2013/ICE-2801, and from the CSIC Research Platform
on Quantum Technologies PTI-001. S.H. is supported by STFC grant ST/T000813/1.

\bibliographystyle{apsrev4-1}
\bibliography{bibliography}

\end{document}